\documentclass[twocolumn]{aastex63}

\usepackage{color}
\usepackage{amsmath}
\usepackage{amssymb}
\usepackage{xcolor}
\usepackage{mathrsfs}
\usepackage{natbib}
\usepackage{gensymb}
\usepackage{xspace}
\usepackage{subfigure}
\usepackage{placeins}

\newcommand{\Rj}{\ensuremath{R_{\rm{Jup}}}\xspace}
\newcommand{\Mj}{\ensuremath{M_{\rm{Jup}}}\xspace}

\newcommand{\Rsun}{\ensuremath{R_{\odot}}\xspace}
\newcommand{\Msun}{\ensuremath{M_{\odot}}\xspace}
\newcommand{\Lsun}{L_\odot}
\newcommand{\Teff}{\ensuremath{T_\mathrm{eff}}\xspace}
\newcommand{\logg}{\ensuremath{\log{g}}\xspace}
\newcommand{\lbollsun}{\ensuremath{\log(L_\mathrm{bol}/\Lsun)}}
\newcommand{\lbol}{\ensuremath{L_\mathrm{bol}}}
\newcommand{\vsini}{\ensuremath{v \sin i}\xspace} 
\newcommand{\angstrom}{\textup{\AA}}
\newcommand{\cratio}{\ensuremath{\mathrm{^{12}C / ^{13}C}\xspace}}
\newcommand{\oratio}{\ensuremath{\mathrm{^{16}O / ^{18}O}\xspace}}
\newcommand{\logco}{\ensuremath{\mathrm{log(^{12}CO / ^{13}CO)}\xspace}}
\newcommand{\logcoo}{\ensuremath{\mathrm{log(C^{16}O / C^{18}O)}\xspace}}
\newcommand{\logho}{\ensuremath{\mathrm{log(H_2^{16}O / H_2^{18}O)}\xspace}}
\newcommand{\ho}{\ensuremath{\mathrm{H_2^{16}O / H_2^{18}O}\xspace}}
\newcommand{\co}{\ensuremath{\mathrm{^{12}CO / ^{13}CO}\xspace}}
\newcommand{\coo}{\ensuremath{\mathrm{C^{16}O / C^{18}O}\xspace}}

\newcommand{\caltech}{Department of Astronomy, California Institute of Technology, Pasadena, CA 91125, USA}
\newcommand{\gps}{Division of Geological \& Planetary Sciences, California Institute of Technology, Pasadena, CA 91125, USA}
\newcommand{\ucsc}{Department of Astronomy \& Astrophysics, University of California, Santa Cruz, CA95064, USA}
\newcommand{\keck}{W. M. Keck Observatory, 65-1120 Mamalahoa Hwy, Kamuela, HI, USA}
\newcommand{\ucla}{Department of Physics \& Astronomy, 430 Portola Plaza, University of California, Los Angeles, CA 90095, USA}
\newcommand{\jpl}{Jet Propulsion Laboratory, California Institute of Technology, 4800 Oak Grove Dr.,Pasadena, CA 91109, USA}
\newcommand{\ucsd}{Center for Astrophysics and Space Sciences, University of California, San Diego, La Jolla, CA 92093}
\newcommand{\ifahonolulu}{Institute for Astronomy, University of Hawai`i, 2680 Woodlawn Drive, Honolulu, HI 96822, USA}
\newcommand{\berkeley}{Department of Astronomy, University of California at Berkeley, CA 94720, USA}

\newcommand{\osu}{Department of Astronomy, The Ohio State University, 100 W 18th Ave, Columbus, OH 43210 USA}

\newcommand{\nmsu}{Department of Astronomy, New Mexico State University, PO Box 30001, MSC 4500, Las Cruces, NM 88003, USA}
\newcommand{\bern}{Center for Space and Habitability, University of Bern, Gesellschaftsstrasse 6, 3012 Bern, Switzerland}
\newcommand{\arizona}{James C. Wyant College of Optical Sciences, University of Arizona, Meinel Building 1630 E. University Blvd., Tucson, AZ. 85721}

\newcommand{\carnegiew}{Earth and Planets Laboratory, Carnegie Institution for Science, Washington, DC, 20015}
\newcommand{\toronto}{David A. Dunlap Institute Department of Astronomy \& Astrophysics, University of Toronto, 50 St. George Street, Toronto, ON M5S 3H4, Canada}
\newcommand{\USQ}{Centre for Astrophysics, University of Southern Queensland, Toowoomba, QLD, Australia}
\newcommand{\upenn}{Department of Physics and Astronomy, University of Pennsylvania, Philadelphia PA 19104
}

\shorttitle{Isotopologue ratios in HIP~55507~AB}
\shortauthors{Xuan et al.}

\begin{document}

\title{Validation of elemental and isotopic abundances in late-M spectral types with the benchmark HIP~55507~AB system}

\correspondingauthor{J. Xuan}
\email{jxuan@astro.caltech.edu}

\author[0000-0002-6618-1137]{Jerry W. Xuan}
\affiliation{\caltech}

\author[0000-0003-0774-6502]{Jason Wang}
\affiliation{Center for Interdisciplinary Exploration and Research in Astrophysics (CIERA) and Department of Physics and Astronomy, Northwestern University, Evanston, IL 60208, USA}

\author[0000-0002-1392-0768]{Luke Finnerty}
\affiliation{\ucla}

\author{Katelyn Horstman}
\affiliation{\caltech}

\author{Simon Grimm}
\affiliation{\bern}

\author[0000-0003-2461-6881]{Anne E. Peck}
\affiliation{\nmsu}

\author[0000-0001-6975-9056]{Eric Nielsen}
\affiliation{\nmsu}

\author[0000-0002-5375-4725]{Heather A. Knutson}
\affiliation{\gps}

\author{Dimitri Mawet}
\affiliation{\caltech}
\affiliation{\jpl}

\author[0000-0002-0531-1073]{Howard Isaacson}
\affiliation{\berkeley}
\affiliation{\USQ}

\author[0000-0001-8638-0320]{Andrew W. Howard}
\affiliation{\caltech}

\author[0000-0003-2232-7664]{Michael C. Liu}
\affiliation{\ifahonolulu}

\author[0000-0001-7062-815X]{Sam Walker}
\affiliation{\ifahonolulu}

\author{Mark W. Phillips}
\affiliation{\ifahonolulu}

\author{Geoffrey A. Blake}
\affiliation{\gps}

\author[0000-0003-2233-4821]{Jean-Baptiste Ruffio}
\affiliation{\caltech}

\author[0000-0003-0097-4414]{Yapeng Zhang}
\affiliation{\caltech}

\author[0000-0001-9164-7966]{Julie Inglis}
\affiliation{\gps}

\author[0000-0003-0354-0187]{Nicole L. Wallack}
\affiliation{\carnegiew}

\author[0000-0002-1838-4757]{Aniket Sanghi}
\affiliation{\caltech}

\author[0000-0002-9329-2190]{Erica J. Gonzales}
\affiliation{\ucsc}

\author[0000-0002-8958-0683]{Fei Dai}
\affiliation{\caltech}
\affiliation{\gps}

\author{Ashley Baker}
\affiliation{\caltech}

\author{Randall Bartos}
\affiliation{\jpl}

\author{Charlotte Z. Bond}
\affiliation{UK Astronomy Technology Centre, Royal Observatory, Edinburgh EH9 3HJ, United Kingdom}

\author[0000-0002-6076-5967]{Marta L. Bryan}
\affiliation{\toronto}

\author[0000-0003-4737-5486]{Benjamin Calvin}
\affiliation{\caltech}
\affiliation{\ucla}

\author{Sylvain Cetre}
\affiliation{\keck}

\author[0000-0001-8953-1008]{Jacques-Robert Delorme}
\affiliation{\keck}

\author{Greg Doppmann}
\affiliation{\keck}

\author{Daniel Echeverri}
\affiliation{\caltech}

\author[0000-0002-0176-8973]{Michael P. Fitzgerald}
\affiliation{\ucla}

\author[0000-0001-5213-6207]{Nemanja Jovanovic}
\affiliation{\caltech}

\author[0000-0002-4934-3042]{Joshua Liberman}
\affiliation{\caltech}
\affiliation{\arizona}

\author[0000-0002-2019-4995]{Ronald A. L\'opez}
\affiliation{\ucla}

\author[0000-0002-0618-5128]{Emily C. Martin}
\affiliation{\ucsc}

\author{Evan Morris}
\affiliation{\ucsc}

\author{Jacklyn Pezzato}
\affiliation{\caltech}

\author[0000-0003-4769-1665]{Garreth Ruane}
\affiliation{\jpl}

\author[0000-0003-1399-3593]{Ben Sappey}
\affiliation{\ucsd}

\author{Tobias Schofield}
\affiliation{\caltech}

\author{Andrew Skemer}
\affiliation{\ucsc}

\author{Taylor Venenciano}
\affiliation{Physics and Astronomy Department, Pomona College, 333 N. College Way, Claremont, CA 91711, USA}

\author[0000-0001-5299-6899]{J. Kent Wallace}
\affiliation{\jpl}

\author[0000-0002-4361-8885]{Ji Wang}
\affiliation{\osu}

\author{Peter Wizinowich}
\affiliation{\keck}

\author[0000-0002-6171-9081]{Yinzi Xin}
\affiliation{\caltech}

\author[0000-0003-2429-5811]{Shubh Agrawal}
\affiliation{\upenn}

\author[0000-0001-5173-2947]{Clarissa R. Do Ó}
\affiliation{\ucsd}

\author[0000-0002-5370-7494]{Chih-Chun Hsu}
\affil{Center for Interdisciplinary Exploration and Research in Astrophysics (CIERA), Northwestern University,
1800 Sherman, Evanston, IL, 60201, USA}

\author[0000-0001-5610-5328]{Caprice L. Phillips}
\affiliation{\osu}

\begin{abstract}
M dwarfs are common host stars to exoplanets but often lack atmospheric abundance measurements. Late-M dwarfs are also good analogs to the youngest substellar companions, which share similar $\Teff\sim2300-2800$ K. We present atmospheric analyses for the M7.5 companion HIP 55507 B and its K6V primary star with Keck/KPIC high-resolution ($R\sim35,000$) $K$ band spectroscopy. First, by including KPIC relative radial velocities between the primary and secondary in the orbit fit, we improve the dynamical mass precision by $60\%$ and find $M_B={88.0}_{-3.2}^{+3.4}~\Mj$, putting HIP~55507~B above the stellar-substellar boundary. We also find that HIP 55507 B orbits its K6V primary star with $a=38^{+4}_{-3}$ AU and $e=0.40\pm0.04$. From atmospheric retrievals of HIP~55507~B, we measure $\rm [C/H]=0.24\pm0.13$, $\rm [O/H]=0.15\pm0.13$, and C/O=$0.67\pm0.04$. Moreover, we strongly detect $^{13}$CO ($7.8\sigma$ significance) and tentatively detect H$_2^{18}$O ($3.7\sigma$ significance) in companion's atmosphere, and measure $\co=98_{-22}^{+28}$ and $\ho=240_{-80}^{+145}$ after accounting for systematic errors. From a simplified retrieval analysis of HIP~55507~A, we measure $\co=79_{-16}^{+21}$ and $\coo=288_{-70}^{+125}$ for the primary star. These results demonstrate that HIP~55507 A and B have consistent \cratio~and \oratio~to the $<1\sigma$ level, as expected for a chemically homogeneous binary system. Given the similar flux ratios and separations between HIP~55507~AB and systems with young, substellar companions, our results open the door to systematically measuring $^{13}$CO and H$_2^{18}$O abundances in the atmospheres of substellar or even planetary-mass companions with similar spectral types.

\end{abstract}

\section{Introduction}
The elemental abundances of exoplanets and substellar companions encode their accretion history, providing valuable insights into planet and star formation mechanisms. It is now well-recognized that measuring abundance ratios besides C/O are crucial for breaking degeneracies and providing a more complete picture of substellar atmospheres \citep[e.g.][]{Cridland2020,Turrini2021, molliere_Interpreting_2022a, Chachan2023} when compared to abundance measurements of their host stars. Recently, isotopologue ratios have also emerged as an observable in substellar atmospheres \citep{Morley2019, Molliere2019}. \citet{zhang_13COrich_2021} measured $\co=31_{-10}^{+17}$ for the young super Jupiter TYC 8998-760-1 b, while \citet{Line2021} reported $\co=10.2–42.6$ for the hot Jupiter WASP-77~Ab. \citet{Finnerty2023} also reported a tentative $^{13}$CO enrichment for WASP-33~b, although higher signal-to-noise (S/N) data is needed to confirm this result. On the other hand, \citet{zhang_12CO_2021} reported $\co=97_{-17}^{+25}$ for an isolated brown dwarf. These results potentially indicate that the varying \cratio~of these objects can be used to constrain their formation histories. However, more analysis and measurements are required to bolster our confidence in these results \citep{Line2021}.

There are abundant measurements of isotopologues in the stellar literature, especially for giant stars. More recently, studies have measured isotopologue ratios in dwarf stars \citep[e.g.][]{Crossfield2019, Botelho2020, Coria2023}, which are thought to better preserve the initial isotopic abundances in their envelopes compared to giant stars, and therefore useful for constraining galactic chemical evolution \citep{Romano2017}. For context, the Sun has $\cratio=93.5\pm3.1$ and $\oratio= 525\pm21$ \citep{Lyons2018}, while the average local interstellar medium values are $\cratio=69\pm6$ and $\oratio=557\pm30$ \citep{Wilson1999}. In circumstellar disks, the relative isotopic abundances can differ from the inherited interstellar medium values due to processes such as self-shielding. For example, \citet{Calahan2022} showed that in certain regions of the inner disk, self-shielding of CO, C$^{18}$O and UV-shielding of H$_2$O can result in an enhanced $\rm{H_{2}^{18}O}$ abundance at the expense of $\rm{C^{18}O}$. In  \citet{zhang_13COrich_2021}, the authors proposed that ices beyond the CO snow line may be $\rm{^{13}CO}$-rich, so if a planet accreted a significant amount of ice beyond the CO snow line it may exhibit a lower \co~value compared to its host star. However, more detailed modeling work is needed to understand the details of isotopic composition and fractionation chemistry in circumstellar disks \citep{Oberg2023}.

In this work, we study the HIP~55507~AB system, which consists of a M7.5 companion that orbits at $\sim$40 AU from its K6V primary star. The M dwarf companion was initially identified from a radial velocity (RV) trend and later confirmed by adaptive optics imaging \citep{gonzales_TRENDS_2020}. Using $K$-band high-resolution ($R\sim35,000$) spectra from Keck/KPIC, we carry out an atmospheric retrieval analysis of HIP~55507~B to measure the C/O, $\rm [C/H]$, \co, and \ho~in its atmosphere. In addition, we analyze the KPIC spectra of the primary star, HIP~55507~A, to measure its \co~and \coo~using a simplified version of the same framework.

From the high-resolution spectra, we also measure the radial velocities (RV) of both stars to compute their relative RV. Relative RV data have been shown to improve orbital constraints for directly imaged companions especially when the other data only sparsely cover the orbital period \citep{Schwarz2016, Ruffio2019, DoO2023}. We include the KPIC relative RVs in orbit fits to measure the companion's orbital parameters and dynamical mass.

This paper is organized as follows. In \S~\ref{sec:hostprop}, we describe the properties of HIP~55507~A, including an estimate of its age. The Keck/HIRES, Keck/NIRC2, and Keck/KPIC observations and data reduction are detailed in \S~\ref{sec:obs_data}. In \S~\ref{sec:massorbit}, we summarize the orbit fits for HIP~55507~B. \S~\ref{sec:spec_retrievals} lays out our spectral analysis framework for both HIP~55507~A and B, including the retrieval setup. \S~\ref{sec:inj_rec} describes the lessons from our of injection-recovery tests for atmospheric retrievals of HIP~55507~B. The main results of our spectral analysis are described in \S~\ref{sec:retrieval_results}, with our conclusions in \S~\ref{sec:conclude}.

\section{Primary Star Properties}\label{sec:hostprop}
HIP~55507~A is a K6V star located at $25.41$ pc with $M=0.67\pm0.02~\Msun$ and $\Teff=4250\pm90$ K \citep{Yee2017, Sebastian2021, Stassun2019, Anders2022}. By comparing the star's Keck/HIRES optical spectra with an empirical spectral library using the \texttt{SpecMatch-Emp} tool \citep{Yee2017}, we obtain $\rm [Fe/H]=-0.02\pm0.09$ for the star.\footnote{The error bar of 0.09 dex comes from the root-mean-square difference between the measured [Fe/H] of stars in the spectral library and their derived [Fe/H] from \texttt{SpecMatch-Emp}. It is the recommended uncertainty to adopt when using \texttt{SpecMatch-Emp} \citep{Yee2017}.} We tabulate the literature properties of HIP~55507~A in Table~\ref{tab:prop}. HIP~55507~A hosts a low-mass companion first detected from RV and direct imaging as part of the TRENDS survey \citep{gonzales_TRENDS_2020}.

\begin{figure*}[t!]
  \centering
    \centering
  \includegraphics[width=.315\linewidth]{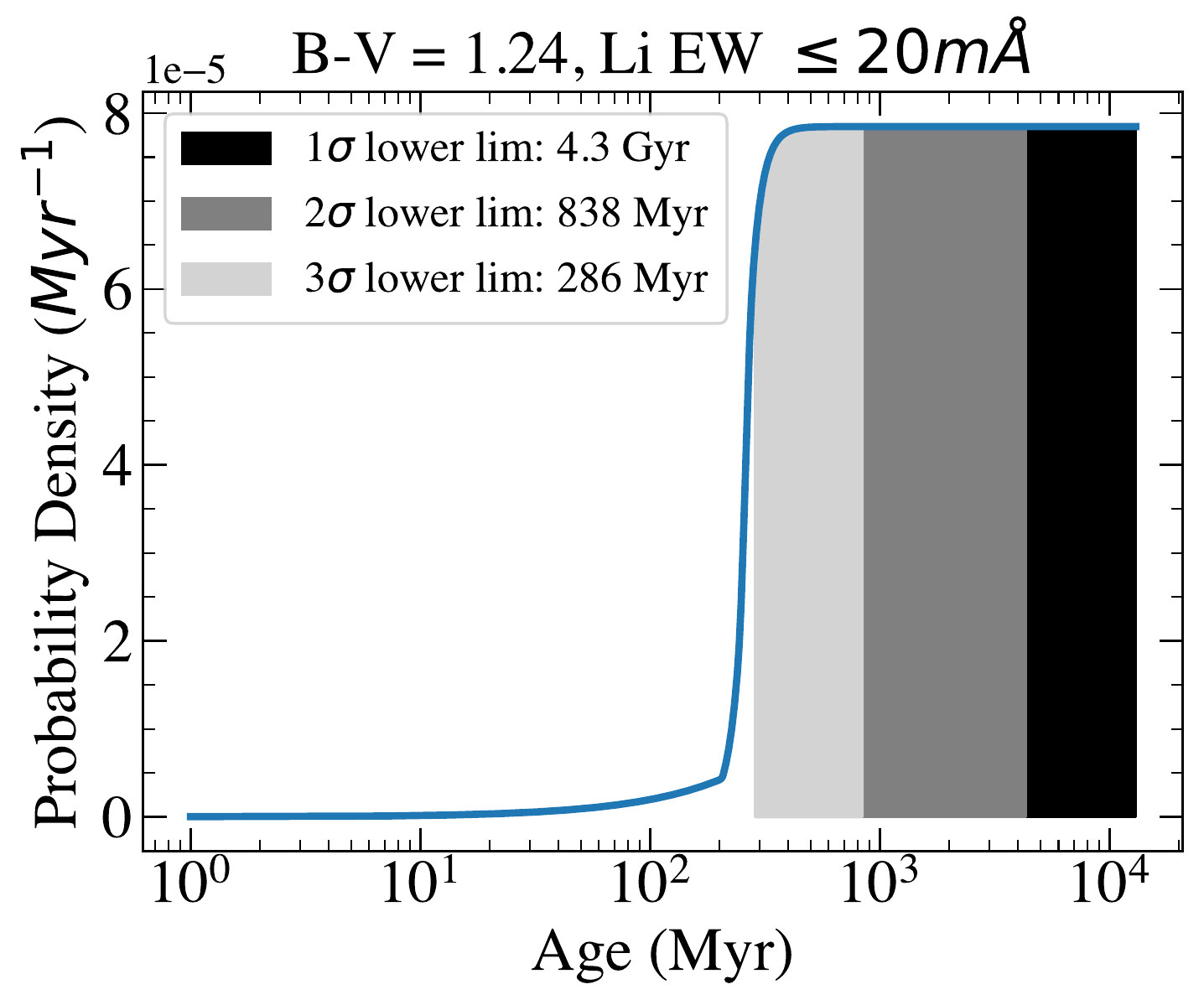}
    \centering
  \includegraphics[width=.345\linewidth]{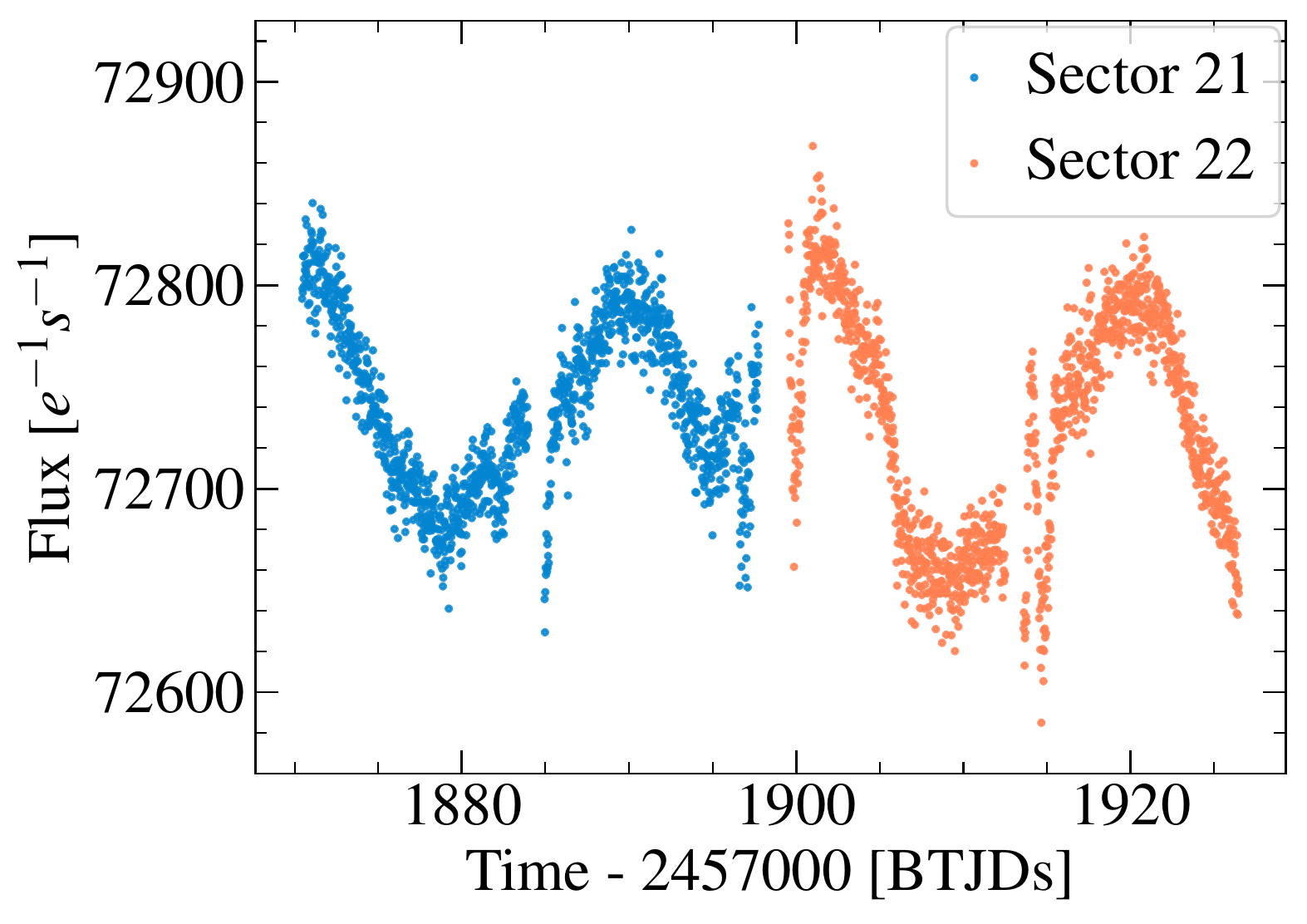}
    \centering
  \includegraphics[width=.325\linewidth]{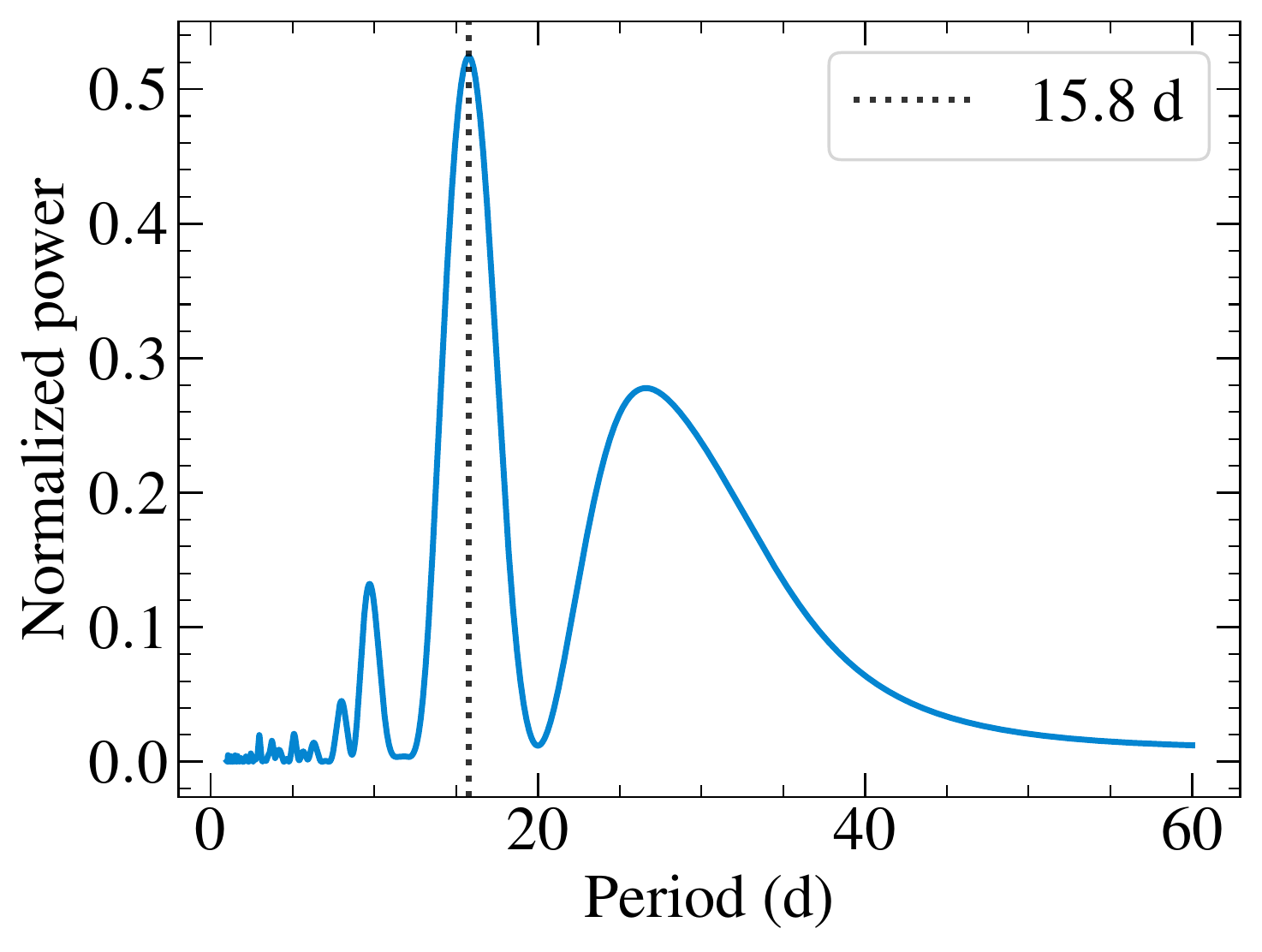}
    \caption{Left: The posterior probability distribution function for the age of HIP 55507 A from BAFFLES (solid blue line), given a lithium equivalent width upper limit of 20 m$\angstrom$ and B-V=1.24. The different shaded regions are 1, 2, and $3\sigma$ lower limits for the age. Middle: TESS light curves from Sectors 21 and 22 extracted from the lightkurve package showing periodic modulation. Right: Interpreting this as rotational modulation, a Lomb-Scargle periodogram shows a rotation period of $15.8\pm1.8$ days. The uncertainties on the rotation period were determined from the full-width half maximum of the peak. These two lines of evidence both point towards an age of $\approx1-2$ Gyr for the star.}\label{fig:age_star}
\end{figure*}

We estimated the age of HIP~55507~A in two ways. First, we searched for lithium with the ARC Echelle Spectrograph \citep{Wang2003} at the Apache Point Observatory 3.5 m on 2023/04/30. The spectrum was reduced with \texttt{pyvista}.\footnote{https://pyvista.readthedocs.io/en/latest/index.html} The spectrum is placed at rest wavelengths by applying a barycentric correction and removing the radial velocity measured by Gaia DR2 \citep{gaia_collaboration_2018}. No Li absorption is visible at 6707.79 $\angstrom$ above the noise and we determine an upper limit of 20 m$\angstrom$ on the lithium equivalent width (EW) by constructing a series of Li lines with Gaussian profiles of varying EWs. With this EW upper limit, we place a lower limit on the stellar age using BAFFLES \citep{StanfordMoore2020}, which uses a Bayesian framework to calculate probability distributions on stellar age for single stars based on Li EW measurements of stars in stellar associations with robust ages. BAFFLES can derive a probability distribution function for a field star given an upper limit on Li EW by using fits to the median Li EW as a function of B-V for each cluster and the scatter about those relations. Given a Li EW upper limit of 20 m$\angstrom$ and B-V=1.24 for HIP~55507~A, we find $2\sigma$ and $3\sigma$ lower age limits of 838 and 286 Myr, respectively (see Fig~\ref{fig:age_star}).

We also searched TESS light curves for rotational modulation using the \texttt{lightkurve} package \citep{lightkurve2018}. HIP~55507~A was observed over two consecutive TESS sectors covering a baseline of 57 days. From the light curves, we found a clear periodic signal of 15.8 days (see Fig~\ref{fig:age_star}). Nearby stars within 15 arcmin do not exhibit similar modulation, suggesting the modulation likely originates from HIP~55507~A. If we attribute the periodic signal to the stellar rotation period, a Lomb-Scargle analysis of the two TESS sectors yields a period of $15.8\pm1.8$ days. Given $\Teff=4250\pm90~$K, we use the gyrochronology tool from \citet{Bouma2023} to derive an age of $1.7^{+0.4}_{-0.7}$ Gyr. Therefore, both the lack of Li and relatively slow rotation point to an age of $\sim1-2$ Gyr for HIP~55507~A.

\begin{deluxetable}{lcc}
\tablecaption{\label{tab:prop}Properties of HIP 55507 AB}
\tablehead{\colhead{Property} & \colhead{Value} & \colhead{References}}
\startdata
\hline
HIP 55507 A \\
$\alpha_{2000.0}$ & 11:22:05.75 & 1\\
$\delta_{2000.0}$ & +46:54:30.2 & 1\\
$\pi$\tablenotemark{a} (mas) & $39.35\pm0.015$ & 1\\
Distance (pc) & $25.41 \pm 0.02$ & 1\\
$\mu_{\alpha}\cos \delta$ (${\rm mas~yr^{-1}}$) & $-197.49 \pm 0.01$ & 1\\
$\mu_{\delta}$ (${\rm mas~yr^{-1}}$) & $-134.78 \pm 0.01$ & 1\\
SpT & K6V & 4\\
Gaia $G$ (mag) & $ 9.271 \pm 0.003 $ & 1\\
$J$ (mag) & $7.367 \pm 0.021$ & 6\\
$H$ (mag) & $6.760 \pm 0.042$ & 6\\
$K_s$ (mag) & $6.613 \pm 0.021$ & 6\\
$W1$ (mag) & $6.544\pm0.075$ & 7\\
$W2$ (mag) & $6.553\pm0.023$ & 7\\
Age (Gyr) & $1.7^{+0.4}_{-0.7}$ & This paper\\
Mass\tablenotemark{b} ($\mathrm{M_\odot}$) & $0.67 \pm 0.02$ & 2,3,4,5\\
Literature $T_{\mathrm{eff}}$ (K) & $4250 \pm 90$ & 2,3,5\\
Literature $\log g$ (dex) & $4.58 \pm 0.06$ & 3,4,5,8\\
Literature $v \sin i$ (${\rm km~s^{-1}}$) & $3.0 \pm 1.0$ & 5\\
$T_{\mathrm{eff}}$ (K) & $4200 \pm 50$ & This paper\\
$\log g$ (dex) & $4.40 \pm 0.25$ & This paper\\
$P_{\rm rot}$ (days) & $15.8 \pm 1.8$ & This paper\\
$\rm [Fe/H]$ & $-0.02\pm0.09$ & This paper\\
\co & $79_{-16}^{+21}$ & This paper\\
\coo & $288_{-70}^{+125}$ & This paper\\
\hline
HIP 55507 B \\
SpT & M7.5 & This paper\\
Mass ($\Mj$) & ${88.0}_{-3.2}^{+3.4}$ & This paper\\
$v \sin i$ (${\rm km~s^{-1}}$) & $5.50 \pm 0.25$ & 
This paper\\
$\rm [C/H]$ & $0.24\pm0.13$ & This paper\\
$\rm [O/H]$ & $0.15\pm0.13$ & This paper\\
C/O & $0.67\pm0.04$ & This paper\\
\co & $98_{-22}^{+28}$  & This paper\\
\ho & $240_{-80}^{+145}$ & This paper\\
\hline
\enddata
\tablenotetext{a}{We correct for the DR3 parallax zeropoint following the guidelines in \citet{Lindegren2021}.}
\tablenotetext{b}{The literature values for the stellar mass, \Teff, and log(g) agree reasonably, so we take the weighted average from the more recent papers and adopt the standard deviation of the different values as the uncertainty for each parameter.}
\tablerefs{(1) \citet{Gaia2023}, (2) \citet{Sebastian2021}, (3) \citet{Stassun2019}, (4) \citet{Petigura_thesis}, (5) \citet{Anders2022}, (6) \citet{cutri_2MASS_2003}, (7) \citet{ALLWISE2013}, (8) \citet{Fouesneau2022}, (9) \citet{Yee2017} }
\end{deluxetable}

\section{Observations and Data Reduction}\label{sec:obs_data}

\subsection{Keck/HIRES}
We collected spectra of HIP~55507~A from April 2009 to June 2023 using the High Resolution Echelle Spectrometer (HIRES, $R\approx60,000$; \citealt{Vogt1994}) at the W.M. Keck Observatory. The data from 2009 to 2015 were collected as part of the M2K program \citep{Gaidos2013}. The observation setup is the same as that used by the California Planet Search \citep{Howard2010}. The wavelength calibration was computed using an iodine gas cell in the light path. A iodine-free template spectrum bracketed by observations of rapidly rotating B-type stars was used to deconvolve the stellar spectrum from the spectrograph point-spread function. We then forward model the spectra taken with the iodine cell using the deconvolved template spectra, the point-spread function model and the iodine cell line atlas \citep{Butler1996}. The Keck/HIRES RVs are presented in Appendix~\ref{app:hires_rvs}, and show a long-term trend with curvature, which is induced by HIP~55507~B (Appendix~\ref{app:nirc2_orbit}).

\subsection{Keck/NIRC2}
We observed HIP 55507 B in $L^\prime$ band on UT 2021 May 19 and $K$ and $M_s$ bands on UT 2022 June 9 using Keck/NIRC2. We did not use a focal plane mask but observed in pupil tracking mode to exploit sky rotation for angular differential imaging (ADI, \citealt{liu_Substructure_2004, marois_adi_2006a}). HIP 55507 B was also imaged with Keck/NIRC2 on UT 2012 Jan 7 and 2015 May 29 (PI: Justin Crepp) as part of the TRENDS survey \citep{gonzales_TRENDS_2020}. The astrometry from \citet{gonzales_TRENDS_2020} shows a $\sim100\degree$ discrepancy in position angle (PA) compared to calibrated images on the Keck Observatory Archive (KOA).\footnote{\url{https://koa.ipac.caltech.edu/cgi-bin/KOA/nph-KOAlogin}}, which could be caused by a mismatch between pupil tracking and field tracking modes used in each observation (E. Gonzales, priv. commun.). Therefore, we re-analyzed the archival NIRC2 data from \citet{gonzales_TRENDS_2020} to update the astrometry. Finally, we include a single astrometric epoch from UT 2021 Dec 21 reported in \citet{Franson2023}.

We first pre-process the data using the Vortex Imaging Processing (VIP) software package \citep{GomezGonzalez2017, Christiaens2023}. We perform flat-fielding, bad-pixel removal, and correct for geometric distortions by applying the solution in \citet{Service2016} for observations after the NIRC2 camera and adaptive optics system were realigned on UT 2015 April 13 and the solution from \citet{Yelda2010} for the archival 2012 observation. Then, we perform sky-subtraction following the procedure described in \citet{xuan_characterizing_2018}. To register the HIP~55507~B frames, we identify the star’s position by fitting a 2D Gaussian to the stellar point spread function (PSF) in each frame.

After obtaining the pre-processed cubes, we extracted the astrometry and photometry of the companion using \texttt{pyKLIP} \citep{wang_pyKLIP_2015}, which models a stellar PSF with Karhunen-Loève Image Processing (KLIP) following the framework in \citet{soummer_Detection_2012b} and \citet{pueyo_DETECTION_2016}. We used ADI to subtract the stellar PSF and tested various model choices to minimize the residuals after stellar PSF subtraction while preserving the companion signal, following guidelines in \citet{Redai2023}. The 2015 observations for HIP~55507 used field tracking mode, so we used a least-squares minimization code to compute the astrometry. We note that our measured astrometry from the archival \citet{gonzales_TRENDS_2020} data agree at the $<1\sigma$ level with those reported in \citet{Franson2023}, who also re-analyzed these data. 

\begin{deluxetable*}{ccccccccc}[t!]
\tablecaption{NIRC2 Astrometry and Photometry for HIP 55507 B} \label{tab:nirc2_results}
\tablehead{\colhead{Time (JD-2400,000)} &  \colhead{UT Date} & \colhead{Filter} & \colhead{Sep. (mas)} & \colhead{PA (deg)} & \colhead{$\Delta m$} & \colhead{$m$} & \colhead{$M_{\rm abs}$} & \colhead{Ref.}} 
\startdata
55934.1 & 2012-01-07$^{\rm (a)}$ & $H$ & $475.6\pm3.0$ & $292.33\pm0.36$ & N/A & N/A & N/A & 1\\
55934.1 & 2012-01-07$^{\rm (a)}$ & $K^\prime$ & $475.6\pm3.0$ & $292.50\pm0.36$ & N/A & N/A & N/A & 1\\
57171.7 & 2015-05-29$^{\rm (a)}$ & $K_{\rm cont}$ & $550.7\pm3.0$ & $274.93\pm0.40$ & N/A & N/A & N/A & 1\\
59353.7 & 2021-05-19 & $L^\prime$ & $732.0\pm3.0$ & $254.75\pm0.24$ & $4.75\pm0.01$ & $11.29\pm0.08$ & $9.27\pm0.08$ & 2\\
59569.7 & 2021-12-21$^{\rm (b)}$ & $K_{\rm s}$ & $748\pm5$ & $254.04\pm0.20$ & N/A & N/A & N/A & 3\\
59739.7 & 2022-06-09 & $M_s$ & $773.1\pm5.0$ & $252.04\pm0.40$ & $4.99\pm0.05$ & $11.54\pm0.06$ & $9.52\pm0.06$ & 2\\
59739.7 & 2022-06-09 & $K$ & $767.0\pm3.1$ & $252.30\pm0.23$ & $5.07\pm0.03$ & $11.66\pm0.04$ & $9.63\pm0.04$ & 2\\
60034.96 & 2023-03-31 & $K$ & $789.9\pm3.0$ & $250.96\pm0.40$ & $5.05\pm0.05$ & $11.64\pm0.04$ & $9.61\pm0.05$ & 2\\
\hline
\enddata
\tablecomments{(a) We re-analyzed the data from these epochs to revise the astrometry. Photometry from these epochs are unreliable due to occultation of the central star by the Lyot Coronagraph. (b) This epoch is from \citet{Franson2023}, who did not quote photometry.}
\tablerefs{(1) \citet{gonzales_TRENDS_2020}, (2) This paper, (3) \citet{Franson2023}}
\end{deluxetable*}

\begin{deluxetable*}{cccccccc}[t!]\label{tab:kpicobs}
    \tablecaption{KPIC Observations of HIP~55507~AB}
    \tablehead{\colhead{UT Date} & \colhead{Target} & \colhead{Exposure Time [min]} & \colhead{Airmass} & \colhead{Throughput} & \colhead{Median S/N per pixel} & \colhead{Science Fibers}}
    \startdata
        2021 July 4 & HIP~55507~A & 2 & 1.4 & $1.1\%$ & 150 & 2,3 \\
        2021 July 4 & HIP~55507~B & 40 & 1.4 & $1.1\%$ & 10.6 & 2,3 \\
        2023 May 2 & HIP~55507~A & 4 & 1.1 & $4.6\%$ & 225 & 2,3 \\
        2023 May 2 & HIP~55507~B & 30 & 1.1 & $4.6\%$ & 22.3 & 2,3 \\
    \enddata
    \label{tab:obs}
    \tablecomments{The throughput is end-to-end throughput measured from top of the atmosphere, and varies with wavelength due to differential atmospheric refraction and the instrumental blaze function. The throughput is computed using the HIP~55507~A spectra using its 2MASS $K_s=6.613$ \citep{cutri_2MASS_2003}. We report the $95\%$ percentile throughput over the $K$ band, averaged over all frames. The median spectral S/N per pixel from 2.29-2.49$~\mu$m is also reported.}
\end{deluxetable*}

From \texttt{pyKLIP} forward modeling \citep{Wang2016}, we obtain the flux ratio between the star and companion for each photometric band, which we convert to apparent and absolute magnitudes. For $L^{\prime}$ and $M_{\rm s}$ bands, we scale the flux ratios to the primary star's $W1$ and $W2$ mag respectively.\footnote{We assume the stars have $L^{\prime}$-$W1$ = 0 and $M_{s}^{\prime}$-$W2$ = 0 as these photometric bands are in the Rayleigh–Jeans tail of the spectral energy distribution for HIP~55507~A.} We convert the 2MASS $K$ into MKO $K$ for HIP 55507 A using color relations in \citet{Leggett2006}, before calculating the MKO $K$ for HIP~55507~B. The measured astrometry and photometry are provided in Table~\ref{tab:nirc2_results}, and an example of the \texttt{pyKLIP} forward modeling is shown in Appendix~\ref{app:nirc2_orbit}.

\subsection{Keck/KPIC}\label{sec:data_red_kpic}
We observed the HIP~55507~AB system with the upgraded Keck/NIRSPEC \citep{martin_overview_2018} using the KPIC fiber injection unit (FIU; \citealt{mawet_Observing_2017, delorme_Keck_2021a, Echeverri2022}) on UT 2021 July 4 and 2023 May 2 (see Table~\ref{tab:kpicobs}). The FIU is located downstream of the Keck II adaptive optics system and is used to inject light from a selected target into one of the single-mode fibers connected to NIRSPEC. We obtained $R\sim35,000$ spectra in $K$ band, which is broken up into nine echelle orders from 1.94-2.49~$\mu$m. The observing strategy is similar to that of \citet{wang_Detection_2021}, except we `nodded' between two fibers to enable background subtraction between adjacent frames. We also acquire short exposures of HIP~55507~A before observing the companion, and spectra of a nearby A0 standard star (HIP 56147) at similar airmass. 

We briefly summarize the KPIC data reduction procedure with the public Python pipeline.\footnote{\url{https://github.com/kpicteam/kpic_pipeline}} For details, see \citet{wang_Detection_2021}. First, we apply nod-subtraction between adjacent frames as the spectral traces of each fiber lands on a different location in the detector. We also remove persistent bad pixels identified from the background frames. Then, we use data from the telluric standard star to fit the trace of each column in the four fibers (two of which contain science data) and nine spectral orders, which give the position and standard deviation of the PSF in the spatial direction at each column. The trace positions and widths are additionally smoothed using a cubic spline to mitigate random noise.

For every frame, we extracted the 1D spectra in each column of each spectral order. To remove residual background light, we subtracted the median of pixels that are at least 5 pixels away from every pixel in each column. Finally, we used optimal extraction \citep{horne_optimal_1986} to sum the flux using weights defined by the 1D Gaussian line-spread function (LSF) profiles calculated from spectra of the telluric star.

For our analysis, we use three spectral orders from 2.29-2.49~$\mu$m, which contain strong absorption lines of CO and H$_2$O from the companion, and CO from the primary star. These orders also have relatively few telluric absorption lines. Note that the three spectral orders have gaps in between them, so the KPIC data effectively cover a range of $\approx0.13~\mu$m after accounting for the gaps.

\begin{deluxetable}{lc}
\tablecaption{Selected parameters from orbit fit} \label{tab:orbit}
\tablehead{\colhead{Parameter} & \colhead{Value}} 
\startdata
$M$ ($\Msun$) & $0.675\pm0.037$ \\
$m$ ($\Mj$) & ${88.0}_{-3.2}^{+3.4}$ \\
$a$ (AU) & ${37.8}_{-2.7}^{+3.5}$ \\
inclination (deg) & $119.3\pm0.7$ \\
ascending node (deg) & $219.9\pm0.7$ \\
period (yr) & ${266}_{-33}^{+44}$ \\
argument of periastron (deg) & ${243.9}_{-5.8}^{+5.3}$ \\
eccentricity & $0.40\pm0.04$ \\
epoch of periastron (JD) & ${2549330}_{-12418}^{+16644}$ \\
\enddata
\tablecomments {A Gaussian prior of $0.67\pm0.04~\Msun$ was imposed on the primary star mass.}
\end{deluxetable}

\begin{figure*}
  \centering
  \begin{subfigure}
    \centering
  \includegraphics[width=.4\linewidth]{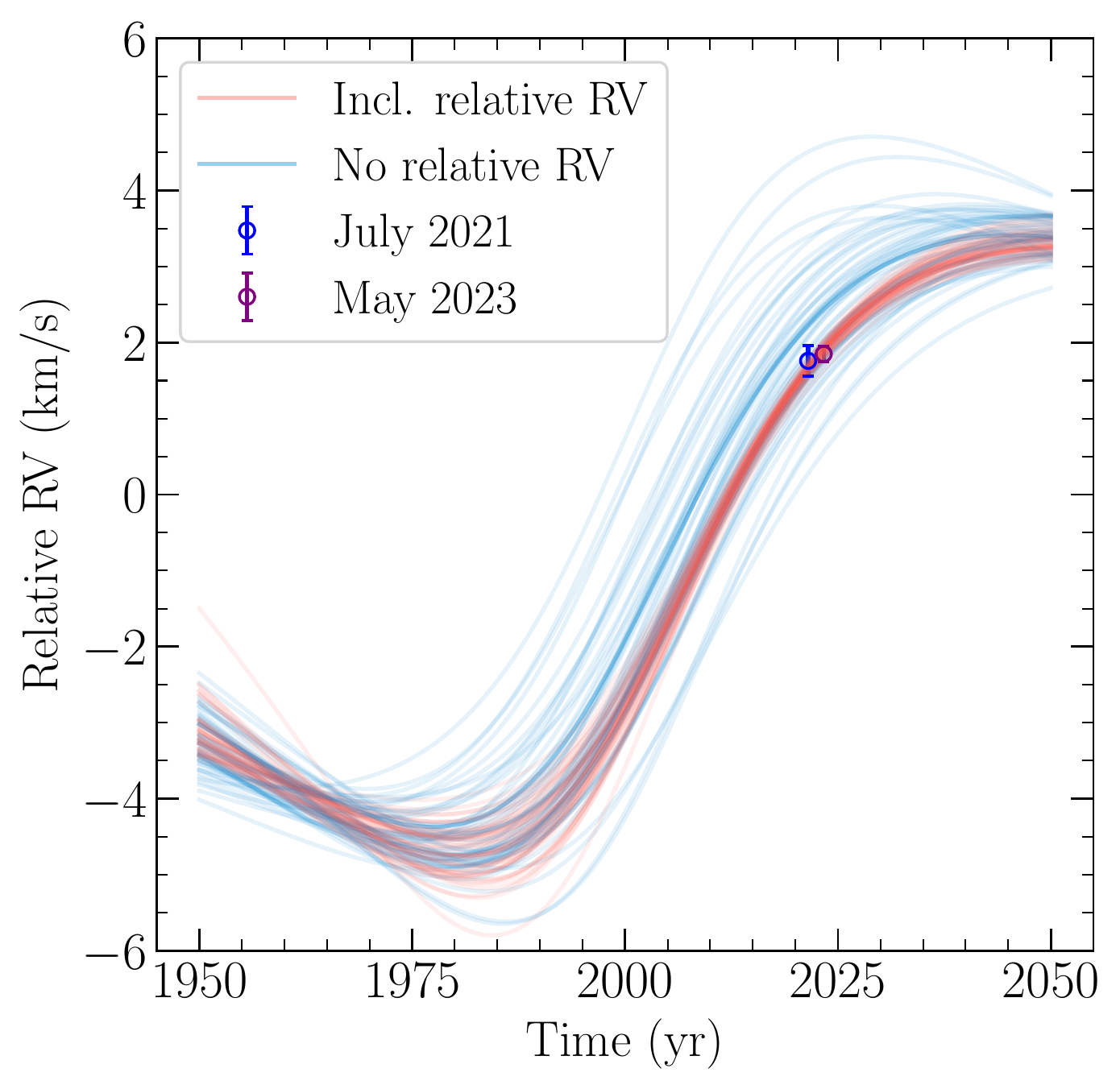}
  \end{subfigure}
  \hspace{10mm}
  \begin{subfigure}
    \centering
  \includegraphics[width=.4\linewidth]{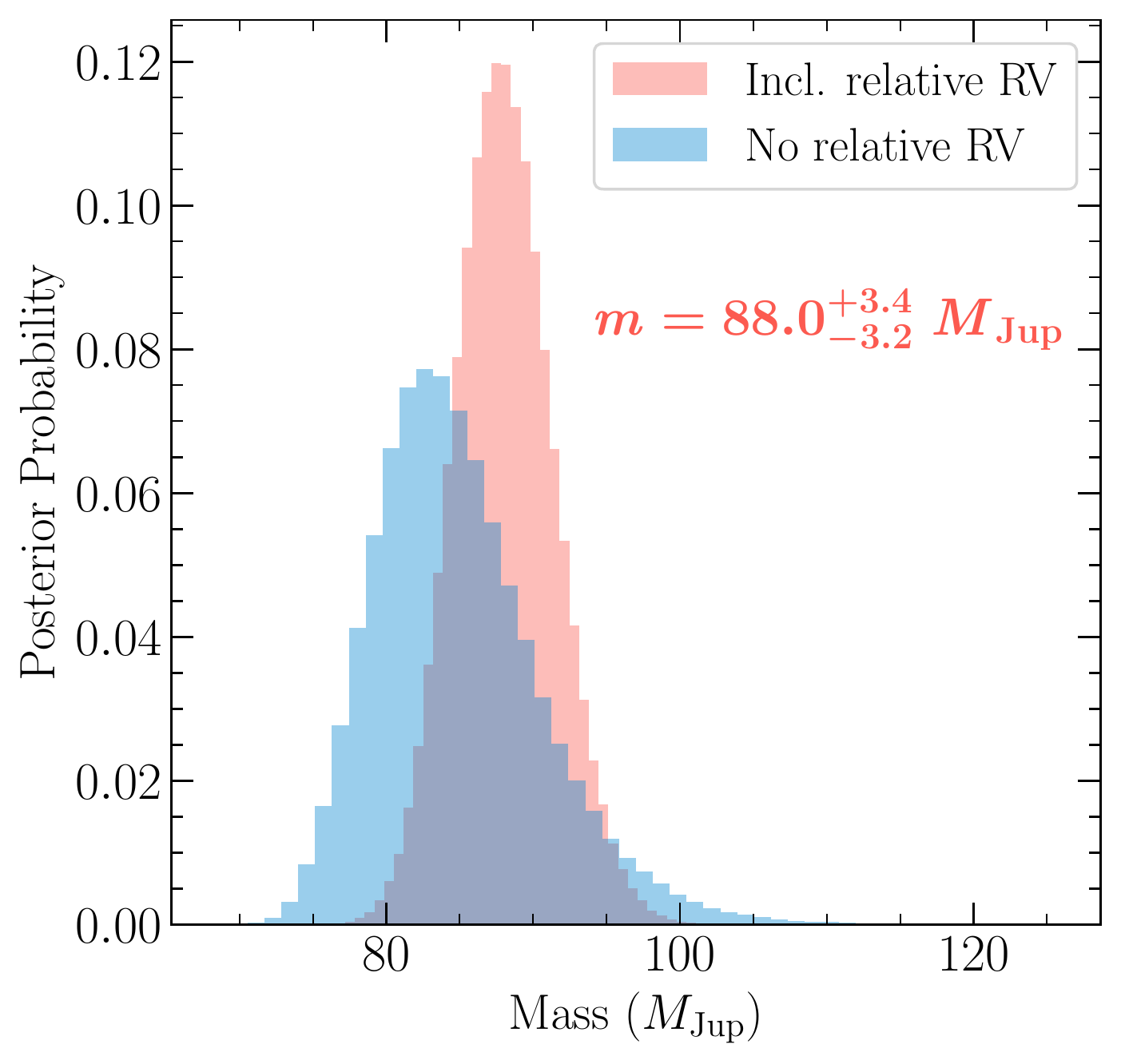}
  \end{subfigure}
    \caption{Left: model relative RV of the HIP~55507~AB system over time from random draws to the posterior distributions of the orbit fits. The red and blue curves are from the fit with and without the relative RVs, respectively. The data points show the observed relative RVs from KPIC. Right: The companion mass posteriors with (red) and without (blue) using the relative RVs, which show a reduction in the mass uncertainty and a slight shift of the median value when the relative RV is incorporated.}\label{fig:relrv}
\end{figure*}

\section{Basic Properties of HIP~55507~B}\label{sec:massorbit}
\subsection{Orbit fits with relative RVs}
The relative radial velocity (RV) between HIP~55507~A and B can be directly measured from our KPIC data. From the two KPIC epochs, we extract two relative RV points (listed in Appendix~\ref{app:relrv_star}) from fitting the KPIC spectra of both HIP~55507~A and B (see \S~\ref{sec:spec_retrievals}).

In the Hipparcos-Gaia Catalog of Accelerations \citep{brandt_HipparcosGaia_2021}, HIP~55507~A shows a significant proper motion anomaly with S/N of $\approx28$ in the Gaia epoch, with an amplitude that is consistent with being induced by HIP~55507~B. We perform orbit fits using RVs of HIP~55507~A from HIRES, relative astrometry from NIRC2 imaging, Gaia and Hipparcos absolute astrometry, and two relative RV points between HIP~55507~A and B from KPIC. We use the \texttt{orvara} package \citep{brandt_orvara_2021} for these fits, which is able to jointly fit the aforementioned data. For the primary mass, we use a Gaussian prior of $0.67\pm0.04~\Msun$, doubling the standard deviation of $0.02~\Msun$ between literature mass measurements (Table~\ref{tab:prop}). We use log-uniform or uniform priors on the other parameters following \citet{brandt_orvara_2021}. We tested orbit fits where we further increase the width of the primary mass prior to $0.67\pm0.08~\Msun$ (or 12\% of the mass), and find the resulting posterior for companion mass shifts by $<1\%$, while the uncertainties on all parameters are consistent to the $<15\%$ level.

Our orbit and mass measurements are summarized in Table~\ref{tab:orbit}. We find a dynamical mass of ${88.0}_{-3.2}^{+3.4}~\Mj$ from this baseline fit. We run a second orbit fit that excluded the KPIC relative RVs to assess their effect on the results. We find that the addition of the two relative RV points from KPIC reduces the companion's mass uncertainty by $\approx60\%$ and shifts the median of the mass posterior to slightly higher values, as shown in Fig.~\ref{fig:relrv}. The uncertainty on the orbital eccentricity also reduces by $\approx50\%$ when including the relative RVs, and we find a moderate eccentricity of $0.40\pm0.04$. In Fig.~\ref{fig:relrv}, we visualize the effect of the relative RVs by plotting random draws from the posteriors of the relative RV (red) and no relative RV fit (blue). While the overall orbital trend is constrained by the other data, the KPIC relative RVs help narrow down the spread in relative RV space, thereby reducing the companion mass uncertainty.

\subsection{Bulk properties and evolutionary state}
\begin{figure}
  \centering
  \includegraphics[width=.8\linewidth]{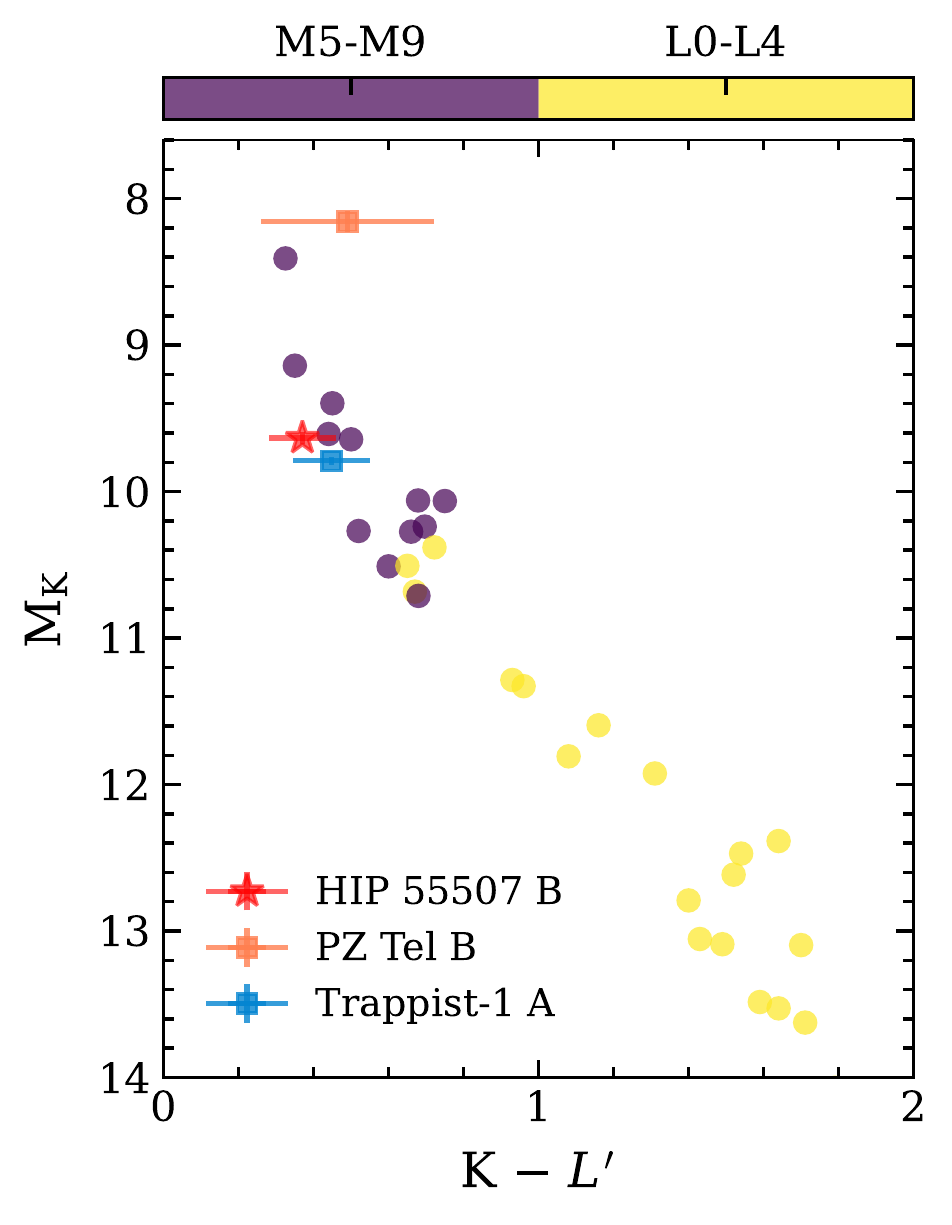}
    \caption{A color-magnitude diagram with $M_K$ and $K-L^{\prime}$. HIP~55507~B is shown as the red star, whereas purple and yellow points in the background are field brown dwarfs with late-M and early-L spectral types, respectively. We also label PZ Tel B, a late-M type substellar companion \citep{Biller2010, Maire2016, Stolker2020}, and Trappist-1 A \citep{cutri_2MASS_2003, ALLWISE2013}, which have properties similar to HIP~55507~B.}
    \label{fig:cmd}
\end{figure}

We place HIP~55507~B on a color-magnitude diagram (CMD) in Fig.~\ref{fig:cmd}. As shown, HIP~55507~B is consistent with a late-M spectral type, and is located very close to Trappist-1~A (an M8.0 star; \citealt{Gillon2016}) on the CMD. Indeed, Trappist-1~A has an inferred mass of $93\pm6~\Mj$ from model fitting \citep{Grootel2018}, very similar to the dynamical mass we measure for~HIP 55507~B. Using relations in \citet{Dupuy2012} and our measured absolute $K_{\rm MKO}$ of $9.63\pm0.04$, we estimate a spectral type of M$7.5\pm0.5$ for HIP~55507~B.

\begin{figure}
  \centering
  \includegraphics[width=\linewidth]{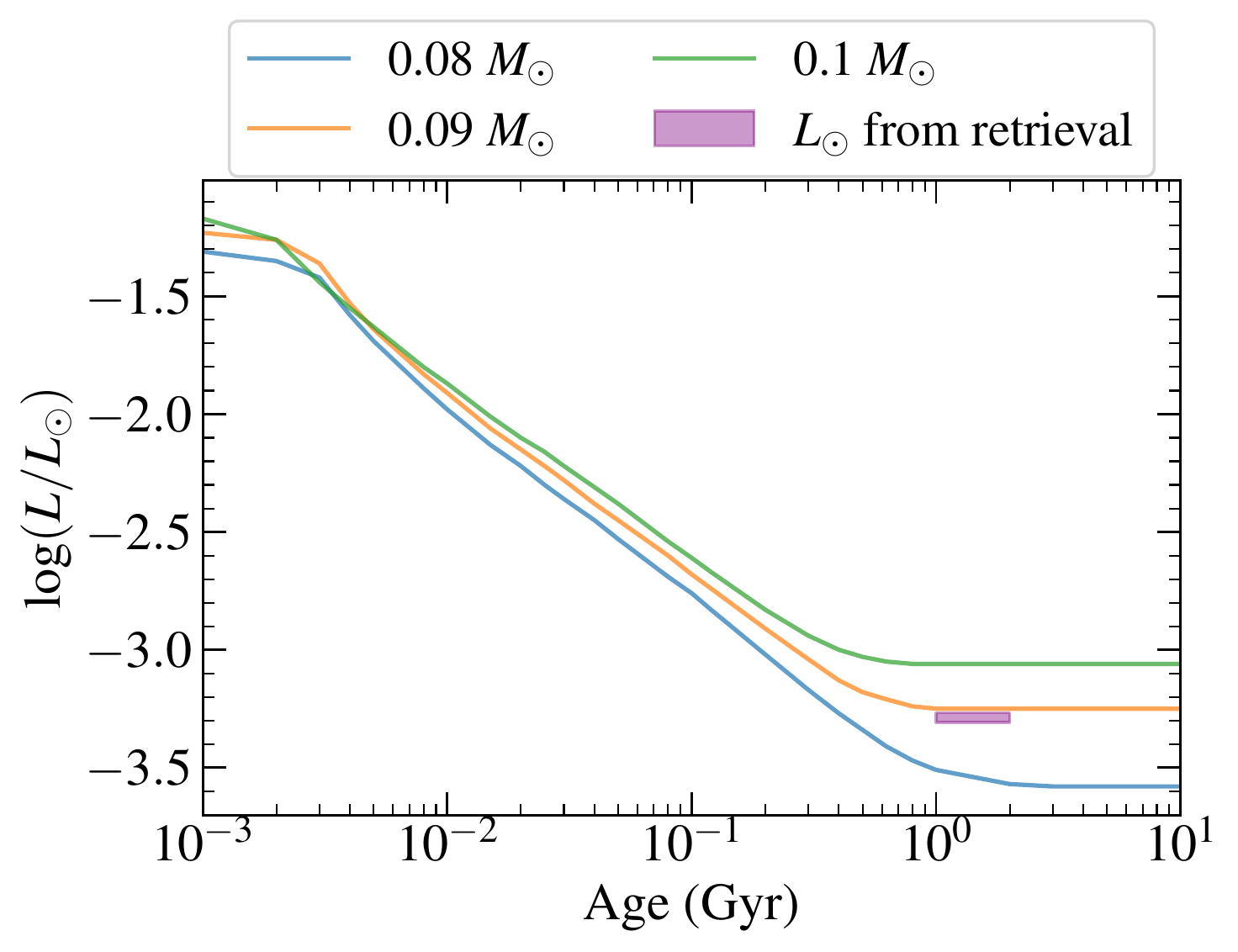}
    \caption{Isochrones from \citet{Baraffe2015} for three different masses. We show the measured \lbol~for HIP~55507~B from spectral retrievals and the estimated age of 1-2 Gyr for the system as the purple shaded region. Both this region and the dynamical mass ($0.084\pm0.003\Msun$) lie in between the $0.08\Msun$ and $0.09\Msun$ isochrones, suggesting that HIP~55507~B's measured properties are consistent with the evolutionary model, and that it is likely on the hydrogen-burning main sequence.}\label{fig:isochrone}
\end{figure}

From our spectral retrievals on the KPIC $K$ band spectra ($R\sim35,000$) and $K$, $L^\prime$, $M_s$ photometry, we estimate $\lbollsun=-3.29\pm0.02$ for HIP~55507~B (see details in \S~\ref{sec:retrieval_results}).\footnote{As a second estimate of \lbol, we use the empirical relation from \citet{Sanghi2023} between $K_{\rm MKO}$ and $L_{\rm bol}$ to obtain $\lbollsun=-3.21\pm0.08$, where the rms scatter of the empirical relation is folded into our luminosity uncertainty. This is consistent with our spectrally-derived \lbol.} In Fig.~\ref{fig:isochrone}, we place HIP~55507~B's \lbol~on isochrone tracks from \citet{Baraffe2015}, and find that the companion falls between the 
$0.08\Msun$ and $0.09\Msun$ isochrones. Therefore, the dynamical mass and \lbol~ for HIP~55507~B are consistent with the \citet{Baraffe2015} substellar model for an age of $\sim1-2$ Gyr, suggesting that the companion has likely reached the hydrogen-burning main-sequence.

\section{Spectral Analysis Framework}\label{sec:spec_retrievals}
In this section, we describe the framework to analyze KPIC spectroscopy of both HIP~55507~A and B. First, we describe the forward model for KPIC (\S~\ref{sec:fm_kpic}), including the model we use to fit fringing modulations (\S~\ref{sec:fringe_model}). Then, we describe the PHOENIX-ACES grid model fits to HIP~55507~A spectra (\S~\ref{sec:A_phoenix}). Lastly, we layout the atmospheric retrieval setup (\S~\ref{sec:retrieval_setup}), which is applied to both HIP~55507~A and B to measure their molecular and/or isotopic abundances.

\subsection{Forward model of the KPIC spectrum}\label{sec:fm_kpic}
Our forward model for KPIC spectra largely follows the framework in \citet{Xuan2022}, with a few updates. In summary, we generate atmospheric templates with \texttt{petitRADTRANS} \citep{molliere_petitRADTRANS_2019, molliere_Retrieving_2020}. These templates are shifted in RV and rotational broadening is performed using the open-source function from \citet{Carvalho_2023}.\footnote{We note that the commonly used \texttt{fastRotBroad} function from \texttt{PyAstronomy} \citep{Czesla_pyA_2019} is only valid for small wavelength arrays, and the question of how small depends on spectral resolution. At $R\sim35,000$, our injection-recovery tests (\S~\ref{sec:inj_rec}) show \texttt{fastRotBroad} can lead to $v\sin{i}$ biases at the $\sim10\%$ level for $v\sin{i}\sim 5$ km/s. In contrast, the \citet{Carvalho_2023} method is accurate over arbitrarily large wavelength grids.} 

Next, we convolve the RV-shifted and rotationally-broadened templates with the instrumental LSF, which we determine from the spectral trace widths in the spatial direction (\S~\ref{sec:data_red_kpic}). As noted by \citet{wang_Detection_2021}, NIRSPEC was designed with a difference in focal lengths in the spatial and dispersion directions by a factor of 1.13 \citep{Robichaud1998}. Following \citet{wang_Detection_2021}, we conservatively allow the LSF width to vary between 1.0 and 1.2 times the width measured in the spatial direction when generating the instrument-convolved companion templates. This uncertainty propagates to our $v\sin{i}$ uncertainty. 

Next, the atmospheric template is multiplied by the telluric and instrumental response, which we determine from spectra of the standard star HIP~56147. For the primary star, HIP~55507~A, that completes the forward model. For the companion, HIP~55507~B, the above procedures constitute a portion of its forward model. The other portion we need to consider for the companion is speckle contribution from the primary star, which we find to account for $\sim1-10\%$ of the total flux in HIP~55507~B's spectra given the relatively small separation of $\approx0.75-0.8\arcsec$ between HIP~55507~A and B. To model the speckle contribution in the companion data, we use observations of HIP~55507~A taken immediately before the companion exposures.

\begin{figure*}[t!]
  \centering
  \begin{subfigure}
    \centering
  \includegraphics[width=.43\linewidth]{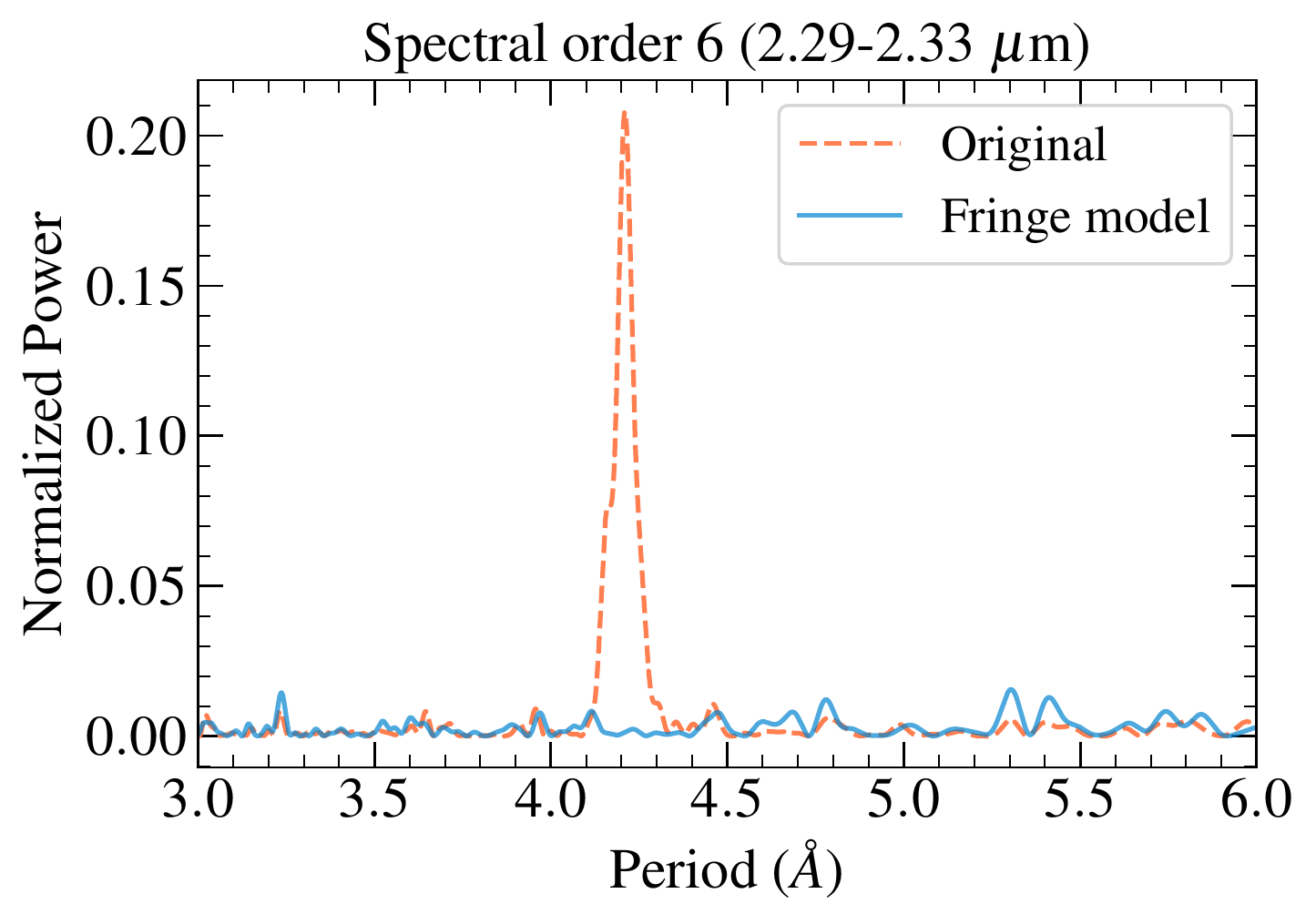}
  \end{subfigure}
  \hspace{5mm}
  \begin{subfigure}
    \centering
  \includegraphics[width=.43\linewidth]{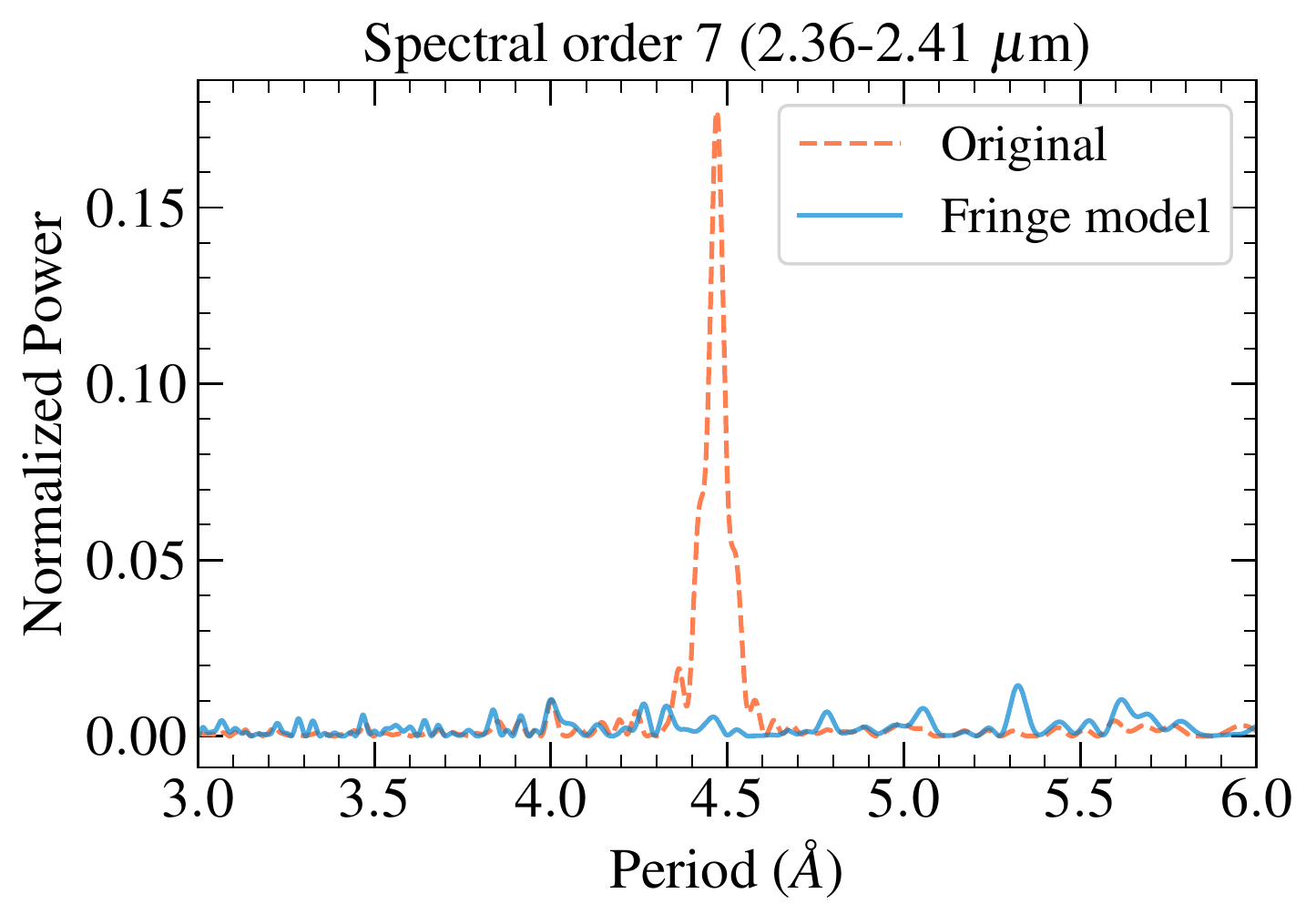}
  \end{subfigure}
    \caption{Lomb-Scargle periodograms of the residuals from fitting the KPIC spectra of HIP~55507~A. The blue and red periodograms are generated from the residuals of fits with and without the fringing model, respectively. The power between $\approx 4-4.5 \AA$, which is the characteristic frequency of fringing from the KPIC dichroic, is greatly diminished. The two panels are for different spectral orders.}\label{fig:fringe_powerA}
\end{figure*}

Finally, we flux-normalize the companion and stellar models and multiply them by different flux scale factors, which are in units of counts as measured by NIRSPEC. After scaling, the companion and speckle models are added in the case of HIP~55507~B. To remove the smoothly varying continuum in the KPIC spectra, we apply high-pass filtering with a median filter of 100 pixels ($\sim0.002~\mu$m) on the data and forward models for both HIP~55507~A and B. The choice of 100 pixels was found to be optimal for KPIC data from \citet{Xuan2022} for accurately retrieving molecular abundances in KPIC data. To summarize, the forward model for HIP~55507~B is:
\begin{equation}
    FM_B = \alpha_c T M_c + \alpha_s D_s
    \label{eqn:comp}
\end{equation}
where $FM_B$ denotes the forward model for HIP~55507~B, $\alpha_c$ and $\alpha_s$ are the flux scales of the companion and speckle, $T$ is the telluric and instrumental response, $M_c$ is the companion template from \texttt{petitRADTRANS}, and $D_s$ is the observed KPIC spectra of HIP~55507~A, which already has $T$ factored in. In contrast, the forward model for HIP~55507~A is:
\begin{equation}
    FM_A = \alpha_A T M_A
     \label{eqn:star}
\end{equation}
where $\alpha_A$ is the flux scale of the primary star in its on-axis observations, $M_A$ is the primary star template from \texttt{petitRADTRANS}, and $T$ is the same transmission function as above. The median filter is applied to both sides of these equations. In reality, we find additional modulation in the HIP~55507~AB data from fringing, which we also account for in our models.

\subsubsection{Fringing model for KPIC data}\label{sec:fringe_model}
KPIC data are affected by a time-varying fringing effect that produces quasi-periodic wiggles in the spectra that can imitate spectral absorption features \citep{Finnerty2022}. Given the high S/N of the HIP~55507~A and B spectra, we noticed the dominant component in the residuals is due to fringing. We describe the details of our fringing model in Appendix~\ref{app:fringeA}. Here, we simply point out that one optic in KPIC (a dichroic) causes the fringing signal to change between the HIP~55507~A spectra and the off-axis HIP~55507~B spectra. The characteristic fringe period induced by this dichroic is $\sim$4$~\AA$ at 2.3~$\mu$m (see Fig.~\ref{fig:fringe_powerA}). We note that the fringe model effectively modifies the $T$ component of Eq.~\ref{eqn:comp} and Eq.~\ref{eqn:star} with an additional transmission term, and therefore applies to all our spectral fits for HIP~55507~A and B.

To incorporate fringing in our spectral fits, we adopt a three-step approach. First, we fit the spectra without the fringe model. The residuals from this first fit are characterized by fringing modulations. Second, we perform a least-squares optimization in the residuals of the first fit to find the best-fit fringing parameters that minimize the fringing signal. Third, in the final spectral fit, we fit the atmospheric parameters and fringe parameters jointly, while adopting the best-fit fringe parameters from the second step as initial guesses. The motivation for this is to avoid the excessively large and complex likelihood space from the fringe model, while also incorporating uncertainties from the fringe model into our atmospheric parameters.

As described in Appendix~\ref{app:fringeA}, our fringe model adds three parameters for each spectral order. In Fig.~\ref{fig:fringe_powerA}, we plot the periodogram of the residuals with and without including the fringing model when fitting HIP~55507~A spectra. The power between $\approx$4-4.5 $\AA$ is noticeably attenuated by the fringe model.

\subsection{PHOENIX-ACES model fits to HIP~55507~A}\label{sec:A_phoenix}
To derive the primary star's bulk properties, we first fit its KPIC spectra with the PHOENIX-ACES model \citep{husser_new_2013}, which here constitutes $M_A$ in Eq.~\ref{eqn:star}. Specifically, we use two spectral orders spanning 2.29-2.41$~\mu$m, with a gap in between. Our model grid assumes solar metallicity, and we vary the \Teff, log(g), RV, \vsini, and stellar flux scale ($\alpha_A$ in Eq.~\ref{eqn:star}). The parameters for the PHOENIX-ACES model fits are summarized in Table~\ref{tab:A_param_prior}. We note that for HIP~55507~A, \vsini acts as a stand-in for the combined effects of rotational broadening and macroturbulence.\footnote{The effects from macroturbulent broadening and rotation are similar and difficult to distinguish at low velocities. Microturbulent broadening is taken into account by the PHOENIX-ACES models, which adopts a microturbulent velocity of $0.48$ km/s \citep{husser_new_2013} for an atmosphere with properties similar to our K6V star ($\Teff=4300~K$, $\logg=4.5$, see Table~\ref{tab:prop}).} The results of the PHOENIX-ACES fits are presented in Appendix~\ref{app:phoenix_A}. Next, we fit the HIP~55507~A spectra using a retrieval framework (see below).

\subsection{Atmospheric retrieval setup}\label{sec:retrieval_setup}
Here, we describe the atmospheric retrieval setup for HIP~55507~B and A to generate $M_c$ and $M_A$ in Eq.~\ref{eqn:comp} and Eq.~\ref{eqn:star}, respectively. Retrievals allow us to measure the isotopologue abundances in both stars. Specifically, we setup radiative transfer routines with \texttt{petitRADTRANS} using the line-by-line opacity sampling method, and down-sample the native $R=10^6$ opacity tables by a factor of 3 to speed up the retrievals. In the following, we describe the opacities, chemistry, temperature profile, and cloud models used in the retrievals. The fitted parameters for HIP~55507~B are summarized in Table~\ref{tab:param_prior}.

We note that compared to the retrieval analysis of HIP~55507~B, our retrievals for HIP~55507~A contain several simplifications, which we highlight throughout this section. Carrying out a retrieval with a free temperature profile and chemical abundances, as we do for HIP~55507~B, is not realistic at this stage for HIP~55507~A. Our $K$ band spectrum for this K6V star is dominated by CO lines with minor contributions from a few atomic lines. H$_2$O is mostly dissociated in the K6V star's photosphere such that we cannot constrain the relative ratios of C and O. With more wavelength coverage (e.g. $H$+$K$ bands to probe OH, CO, CN), a spectral synthesis approach could be a way to measure elemental abundances and C/O for HIP~55507~A, as achieved for a K7V dwarf by \citet{Hejazi2023}.

\subsubsection{Opacity sources}
We require high temperature opacities for our stars. For HIP~55507~B, our preliminary retrievals show that there is contribution to the emission spectrum from regimes with $T>3000~$K (see Fig.~\ref{fig:ptprofile}), which exceeds the $3000~$K upper limit of default \texttt{petitRADTRANS} opacity tables for molecules \citep{molliere_petitRADTRANS_2019}. Therefore, whenever possible, we update our opacity tables to go to $T_{\rm max}=4500~$K or higher using the DACE opacity database generated with the HELIOS-K opacity calculator \citep{Grimm2015, Grimm2021}.\footnote{\url{https://dace.unige.ch/opacityDatabase/}} In particular, we upgrade the line opacities of H$_2^{16}$O \citep{Polyansky2018}, OH \citep{Brooke2016}, FeH \citep{Dulick2003, Bernath2020}, TiO \citep{McKemmish2019}, AlH \citep{Yurchenko2018}, and VO \citep{McKemmish2016} to reach $4500$ K. For H$_2^{18}$O, we adopt the line list from \citet{Polyansky2017} which is valid up to $3000~$K. For H$_2$S, we use the line list from \citet{Azzam2016}, valid up to $2000~$K. Finally, we include the atomic line species Na, K, Mg, Ca, Ti, Fe, and Al \citep{Kurucz2011}.

For HIP~55507~A, the photosphere is at even higher temperatures, but the $K$ band spectra of this star is dominated by mostly CO and the aforementioned atomic lines. Therefore, we only include these opacities for the HIP~55507~A retrievals. We generate opacities for C$^{16}$O, $^{13}$CO, C$^{18}$O that are valid up to 9000~K from \citet{Rothman2010}.

For the continuum opacities in both stars, we include the collision induced absorption (CIA) from H$_2$-H$_2$ and H$_2$-He, as well as the H- bound-free and free-free opacity.

\begin{deluxetable*}{llll}[t!]
\tablecaption{Fitted parameters and priors in HIP~55507~B retrievals}
\label{tab:param_prior}
\tablehead{Parameter & Prior & Parameter & Prior}
\startdata
Mass ($\Mj$) & $\mathcal{N}(88.0, 3.4)$ & Radius ($\Rj$) & $\mathcal{U}(0.6, 2.5)$  \\
$T_{\rm anchor}$ ($K$) & $\mathcal{U}(1900, 2700)$ & RV (km/s) & $\mathcal{U}(-30 , 30)$ \\
$\Delta T_1$ ($K$) & $\mathcal{U}(400, 1000)$ & $v\sin{i}$ (km/s) & $\mathcal{U}(0, 30)$ \\
$\Delta T_2$ ($K$) & $\mathcal{U}(50, 700)$ & $\rm C/O$ & $\mathcal{U}(0.1,1.0)$  \\
$\Delta T_3$ ($K$) & $\mathcal{U}(50, 600)$ & $\rm [C/H]$ & 
$\mathcal{U}(-1.5,1.5)$ \\
$\Delta T_4$ ($K$) & $\mathcal{U}(50, 600)$ & \logco & $\mathcal{U}(0, 6)$ \\
$\Delta T_5$ ($K$) & $\mathcal{U}(50, 600)$ & \logho & $\mathcal{U}(0, 6)$  \\
$\Delta T_6$ ($K$) & $\mathcal{U}(50, 600)$  & \logcoo & $\mathcal{U}(0, 6)$ \\ 
$\Delta T_7$ ($K$) & $\mathcal{U}(50, 600)$  & log(gray opacity/$\rm{cm}^2 g^{-1}$)$^{\rm (a)}$ & $\mathcal{U}(-6, 6)$ \\ 
$f_{\rm sed}^{\rm (b)}$ & $\mathcal{U}(0, 10)$ & ${\rm log}(K_{\rm zz}/\rm{cm^2 s^{-1}})^{\rm (b)}$ & $\mathcal{U}(5, 13)$ \\
$\sigma_{\rm g}^{\rm (b)}$ & $\mathcal{U}(1.05, 3)$ & ${\rm log}(\tilde{X}_{\rm Al_2O_3})^{\rm (b)}$ & $\mathcal{U}(-2.3, 1)$ \\
\hline
Fringe and other parameters &  &  & \\ 
\hline
Optical path length, $d$ (mm) & $\mathcal{U}(4, 5)$ & Comp. flux, $\alpha_c$ (counts) & $\mathcal{U}(0, 300)$ \\
Fractional amplitude, $F$ & $\mathcal{U}(10^{-6}, 1)$ & Speckle flux, $\alpha_s$ (counts) & $\mathcal{U}(0, 200)$  \\ 
Dichroic temperature, $T_d$ ($K$) & $\mathcal{U}(150, 330)$ & Error multiple$^{\rm (c)}$ & $\mathcal{U}(1, 5)$ \\ \hline
\enddata
\tablecomments{$\mathcal{U}$ stands for a uniform distribution, with two numbers representing the lower and upper boundaries. $\mathcal{N}$ stands for a Gaussian distribution, with numbers representing the median and standard deviation. The fringe parameters $d$, $F$, and $T_d$ are described in Appendix~\ref{app:fringeA}.}
\tablenotetext{a}{Parameter for the gray cloud model (constant gray opacity).}
\tablenotetext{b}{Parameters for the EddySed cloud model. $\tilde{X}_{\rm Al_2O_3}$ is the scaling factor for the cloud mass fraction, so that $\rm{log} (\tilde{X}_{\rm Al_2O_3})=0$ refers to a fraction equal to the equilibrium mass fraction. $f_{\rm sed}$, $K_{\rm zz}$, and $\sigma_{\rm g}$ together set the cloud mass fraction as a function of pressure and the cloud particle size distribution \citep{molliere_Retrieving_2020}.}
\tablenotetext{c}{The error multiple term is fitted for KPIC data to account for any underestimation in the uncertainties.}
\end{deluxetable*}

\begin{deluxetable*}{llll}
\tablecaption{Fitted parameters and priors in HIP~55507~A retrieval and PHOENIX-ACES fit}
\label{tab:A_param_prior}
\tablehead{Parameter (retrieval) & Prior & Parameter (PHOENIX-ACES) & Prior}
\startdata
Mass ($\Msun$) & $\mathcal{U}(0.63, 0.71)$ & \Teff ($K$) & $\mathcal{U}(3000, 5500)$ \\
Radius ($\Rsun$) & $\mathcal{U}(0.64, 0.72)$ & \logg (dex) & $\mathcal{U}(3.5, 5.5)$  \\
$v\sin{i}$ (km/s) & $\mathcal{U}(0, 30)$ & $v\sin{i}$ (km/s) & $\mathcal{U}(0, 30)$ \\
RV (km/s) & $\mathcal{U}(-30 , 30)$ & RV (km/s) & $\mathcal{U}(-30 , 30)$ \\
$\rm C/O$ & $\mathcal{N}(B_\mu, B_\sigma)$ &   \\
$\rm [C/H]$ & $\mathcal{N}(B_\mu, B_\sigma)$ & \\
\logco & $\mathcal{U}(0, 6)$ & \\
\logcoo & $\mathcal{U}(0, 6)$ & \\
\hline
Fringe and other parameters &  &  & \\ 
\hline
Optical path length, $d$ (mm) & $\mathcal{U}(4, 5)$ & Star flux, $\alpha_A$ (counts) & $\mathcal{U}(0, 10000)$ \\
Fractional amplitude, $F$ & $\mathcal{U}(10^{-6}, 1)$ & Error multiple & $\mathcal{U}(1 , 5)$ \\ 
Dichroic temperature, $T_d$ ($K$) & $\mathcal{U}(150, 330)$ & \\ \hline
\enddata
\tablecomments{Symbols for priors are same as Table~\ref{tab:param_prior}. The parameters for the HIP~55507~A retrieval and PHOENIX-ACES fits are in the left and right columns, respectively. `Fringe and other parameters' are common to both. For C/O and [C/H], $B_\mu$ and $B_\sigma$ represent the median and $1\sigma$ interval measured for the companion, HIP~55507~B, which are used as Gaussian priors in the HIP~55507~A retrieval.}
\end{deluxetable*}

\subsubsection{Chemistry and isotopologue ratios}
The default chemical equilibrium grid in \texttt{petitRADTRANS} does not save the abundances of all the species we include as opacity sources. Therefore, we generate a custom chemical equilibrium grid using \texttt{easyCHEM},\footnote{\url{https://easychem.readthedocs.io/en/latest/content/installation.html}} which is the same code used by \citet{molliere_Retrieving_2020}. We validated our new grid against the \texttt{petitRADTRANS} chemical grid for overlapping species and find excellent agreement (fractional differences of $<1\%$). The abundances of species are set by two parameters in our grid: the carbon abundance $\rm [C/H]$, and the carbon-to-oxygen ratio C/O, which determines the oxygen abundance along with $\rm [C/H]$. We are only sensitive to the abundances of C and O in this work, and therefore assume that the other metals scale with C. Our grid goes up to 8000~$K$, more than hot enough for the K6V primary star. For the solar elemental abundances, we adopt \citet{asplund_Chemical_2009}.

In our retrievals, the abundances of the main isotopologues are obtained from interpolating the chemical equilibrium grid for each value of C/O and $\rm [C/H]$. Then, for each minor isotopologue included, we fit for an isotopologue ratio parameter akin to \citet{zhang_13COrich_2021}. In our baseline retrievals, we fit for three ratios: $\rm {^{12}C^{16}O/^{13}C^{16}O}$, $\rm{^{12}C^{16}O/^{12}C^{18}O}$, and $\rm{H_{2}^{16}O/H_{2}^{18}O}$. This allows us to explore whether the $^{16}$O/$^{18}$O ratio differs between CO and H$_2$O, which may arise from isotopic fractionation processes such as self-shielding of CO and UV shielding of H$_2$O \citep{Calahan2022}. In addition, fitting two ratios allows us to examine whether the data show evidence for $\rm{H_{2}^{18}O}$, $\rm{^{12}C^{18}O}$, or both. 

\subsubsection{Temperature structure}
For HIP~55507~B, we adopt a modified version of the pressure-temperature (P-T) profile from \citet{Piette2020}. Our profile is parameterized by seven $\Delta T/ \Delta P$ values between eight pressure points and the temperature at one of these pressures. Instead of having the pressure points equally spaced in log pressure, we preferentially concentrate pressure points around the peak of the weighted emission contribution, as this is where the data are most informative. The selected pressure points are labeled in Fig.~\ref{fig:ptprofile}, and the modeled pressure extent is between log(bar) = -4.0 to 1.4. For the radiative transfer, the eight P-T points from our profile are interpolated onto a finer grid of 100 P-T points using a monotonic cubic interpolation as recommended by \citet{Piette2020}. Unlike \citet{Piette2020}, we do not apply smoothing to our profiles as smoothing has been shown to bias retrieval results \citep{Rowland2023}.

For HIP~55507~A, we fix its P-T profile to a Phoenix P-T profile \citep{husser_new_2013} matching properties of the star ($\Teff=4300~K$ and log(g)=4.5) for simplicity.

\subsubsection{Clouds}
Clouds are expected to play a minimal role for late-M objects like HIP~55507~B, and no role for a K6V star like HIP~55507~A, as temperatures are too hot for cloud condensates to remain stable. For completeness in our HIP~55507~B retrievals, we consider both clear and cloudy models to explore the sensitivity of our retrieved abundances to assumed cloud properties. For the cloudy models, we use a gray cloud model where a constant opacity is added to the atmosphere, and the EddySed model \citep{ackerman_Precipitating_2001} as implemented in \texttt{petitRADTRANS} \citep{molliere_Retrieving_2020}. We used Al$_2$O$_3$ clouds in the EddySed model, as Al$_2$O$_3$ is expected to be more important at higher \Teff~\citep{wakeford_Transmission_2015}. 

\subsubsection{Summary of retrieval setup for both stars}
As noted earlier, we make simplifications in retrieving the spectra of HIP~55507~A. To summarize the retrieval setup for A: 1)~we adopt priors on C/O and $\rm [C/H]$ for HIP~55507~A using measured values from HIP~55507~B, 2) fix the P-T profile to a Phoenix P-T profile \citep{husser_new_2013}, and, 3) allow the stellar mass and radius to vary within $1\sigma$ intervals given in Table~\ref{tab:prop}.

In contrast, we freely fit for all these parameters in the HIP~55507~B retrievals, with the exception of mass. For the companion's mass, we adopt a dynamical mass prior determined from \S~\ref{sec:massorbit}. The fitted parameters and adopted priors in HIP~55507~B and A retrievals are listed in Table~\ref{tab:param_prior} and Table~\ref{tab:A_param_prior}, respectively.

\subsubsection{Jointly fitting photometry for HIP~55507~B}\label{sec:phot}
For HIP~55507~B, we jointly fit the KPIC high-resolution spectra with the $K$, $L^\prime$, and $M_s$ photometric points in Table~\ref{tab:nirc2_results} to better constrain the bulk properties of the companion. The apparent magnitudes were converted into flux density units by computing the zeropoint for each photometric filter using the \texttt{species} package \citep{Stolker2020}. For the photometry model, we use the correlated-$k$ opacities in \texttt{petitRADTRANS} (re-binned to $R=50$). The photometry model does not add any new parameters, as it is fully described by the atmospheric parameters introduced earlier. In joint KPIC+photometry retrievals, we add the log likelihoods from the photometry and KPIC components.

\subsubsection{Model fitting with nested sampling} \label{sec:ns_bayes}
We use nested sampling as implemented by \texttt{dynesty} \citep{speagle_DYNESTY_2020} to find the posterior distributions for all model parameters listed in Table~\ref{tab:param_prior} for HIP~55507~B, and Table~\ref{tab:A_param_prior} for HIP~55507~A. We use 600 live points and adopt the stopping criterion that the estimated contribution of the remaining prior volume to the total evidence is less than 1\%. We confirmed that increasing the number of live points to 1000 does not meaningfully change the posteriors of our retrieved parameters.

\begin{deluxetable*}{lcccccccc}
\tabletypesize{\footnotesize}
\tablecaption{Input and retrieved parameters from injection-recovery tests}
\tablehead{
Parameter &  RV (km/s) & \vsini (km/s) & C/O & $\rm [C/H]$ & \logco & \logho & \logcoo
}
\startdata
\hline
Input values & 10.0; -10.0 & 5.00 & 0.70 & 0.0 & 2.00 & 2.30 & 2.78 \\
\hline
RV = 10 km/s & $9.96\pm0.03$ & $5.02\pm0.09$ & $0.706\pm0.005$ & $0.02\pm0.03$ & $2.05\pm0.04$ & $2.27\pm0.05$ & ... \\
RV = -10 km/s & $-10.05\pm0.02$ & $4.99\pm0.09$ & $0.693\pm0.005$ & $0.01\pm0.03$ & $1.90\pm0.03$ & $2.33_{-0.11}^{+0.14}$ & ... \\
\hline
\enddata
\tablecomments{The injections were performed on a non-illuminated KPIC fiber to sample realistic thermal noise properties. We place a Gaussian prior on the mass, so do not report this value. \label{table:inj_rec}}
\end{deluxetable*}

\section{Injection-recovery tests}\label{sec:inj_rec}
To validate our retrieval framework, we perform a series of injection-recovery retrievals on simulated data for HIP 55507 B. We inject model spectrum from \texttt{petitRADTRANS} in the extracted spectrum of a non-illuminated fiber. Such a fiber still measures the thermal background of the instrument, and serves as a realistic `noise spectrum' since thermal noise is the dominant source of noise for our data. 
Specifically, we use the extracted fiber 4 trace at the time of observations. We took the fiber 4 trace on exposures where HIP 55507 B was aligned to fiber 2, at which time fiber 4 was located $\sim 2\arcsec$ away from the companion and $\sim 2.5\arcsec$ away from HIP 55507 A. Examination of the extracted spectra from fiber 4 shows that there is negligible leaked light: the median of the flux is $\approx0$ counts. We multiply the companion model by the telluric response function ($T$) for fiber 4, add a speckle flux contribution using the primary star spectra, and high-pass filter the simulated data in the same way as for the real data. 

For the input companion model, we use a P-T profile from the SPHINX model grid \citep{Iyer2023} with $\Teff=2500$ K, log(g)=5.25, solar metallicity, and C/O=0.7. We set the mass ($87~\Mj$) and radius ($1.1~\Rj$) of our injected companion to be consistent with this \logg, and the chemical abundances to C/O=0.7 and $\rm [C/H]=0$, consistent with that assumed in the P-T profile. To achieve a similar S/N as the real data, we inject similar companion and speckle flux values as the data. We carry out two injections at different RV shifts (-10 and 10 km/s) to sample various parts of the background trace. In addition to simulating a KPIC model, we also simulate three photometry points using the same input values. The parameters of the simulated models are given in Table~\ref{table:inj_rec}. 

In our simulated model we inject $\co=100$ ($\logco=2.0$) and $\ho=200$ ($\logho\approx2.301$), similar to what we find in the real data. Since we do not detect \coo~in the data, we inject a lower value of $\coo=600$ ($\logcoo\approx2.778$) to see whether this can be recovered in our tests. Note that our isotopologue ratios are fitted in log scale in the retrieval.

\begin{figure*}[t!]
    \centering
 \includegraphics[width=.48\linewidth]{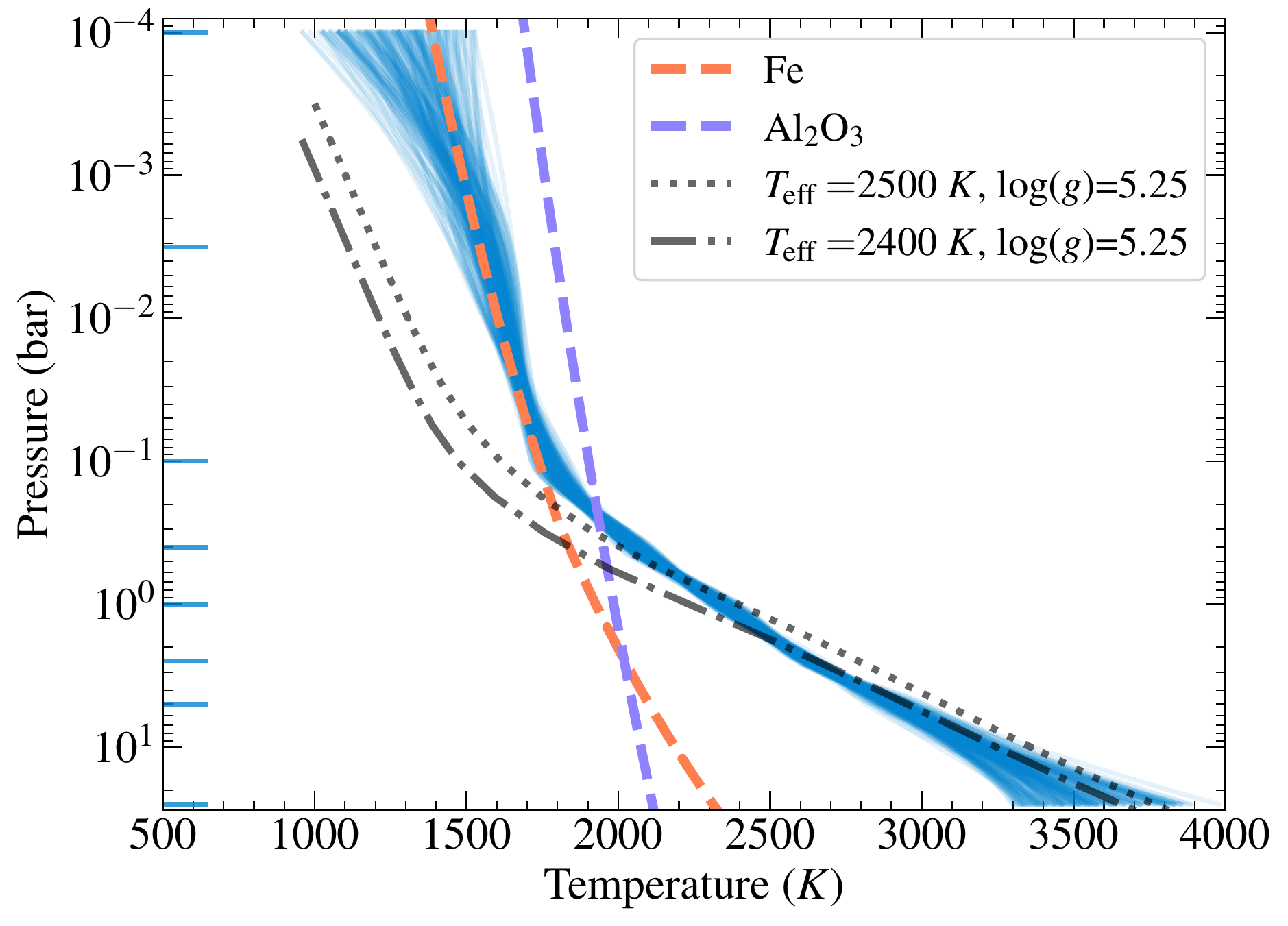}
    \centering
  \includegraphics[width=.46\linewidth]{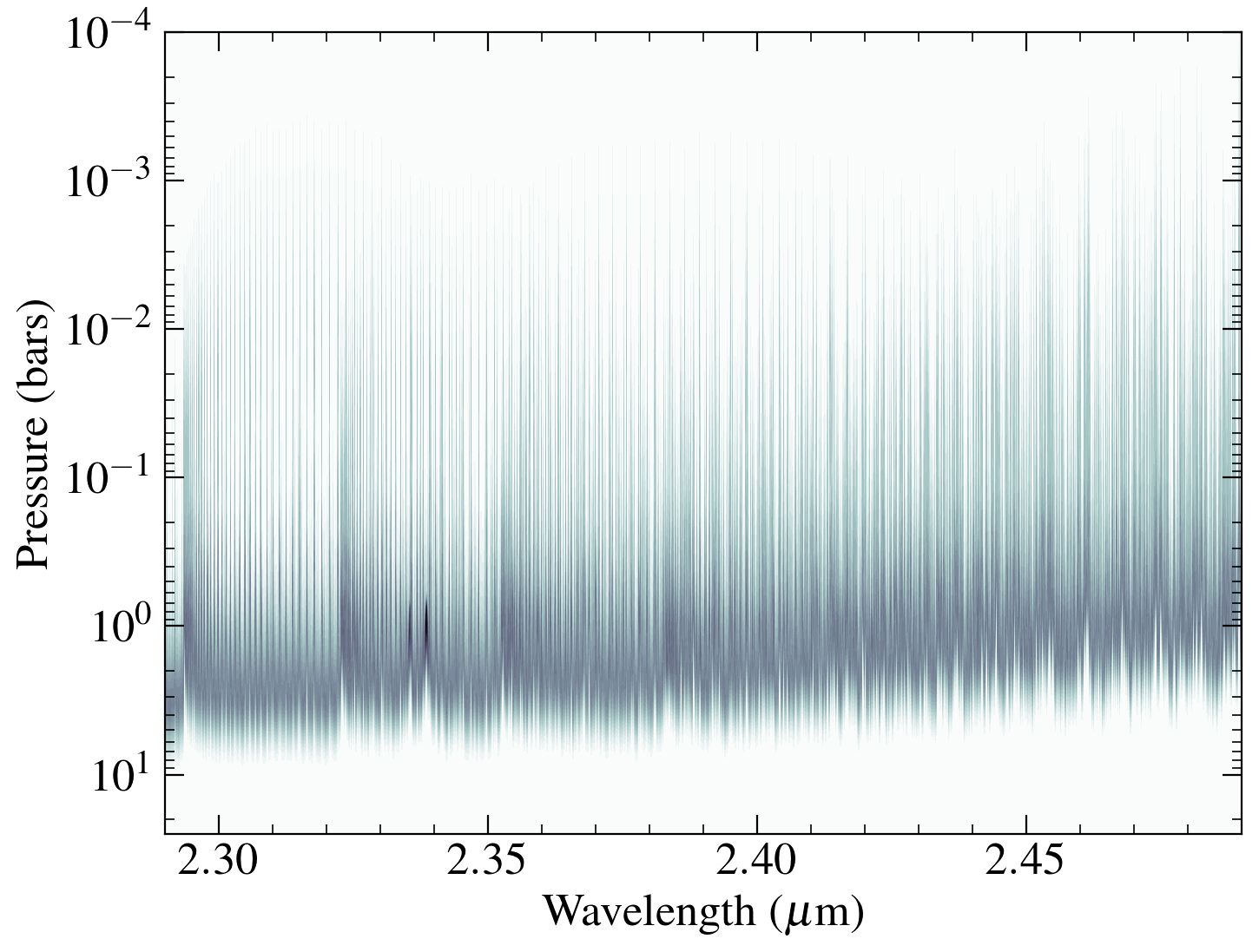}
    \caption{Left: blue: random draws from the posterior of the retrieved P-T profiles from the baseline retrieval. The gray lines show SPHINX models \citep{Iyer2023} with similar bulk properties as HIP~55507~B. The condensation curves for Fe and Al$_2$O$_3$ clouds are plotted in colored dashed lines. The horizontal ticks on the y-axis are pressure points between which we fit $\Delta T$ values in our P-T parameterization. Right: The emission contribution function of the best-fit baseline model. There is non-zero contribution at $\sim 5$ bars, where the temperature profile (left panel) exceeds 3000~$K$.}
    \label{fig:ptprofile}
\end{figure*}

As shown in Table~\ref{table:inj_rec}, \logco~and \logho~are recovered, though systematic offsets of $0.05-0.1$ dex can be present in the retrieved $\logco$. On the other hand, the systematic bias is only $\sim0.03$ dex for \logho. To be conservative, we adopt 0.1 dex as the systematic error for both of these isotopologue ratios in our retrievals of the real data. The \logcoo~posterior is not well-bounded, though the $3\sigma$ lower limit for \logcoo~does contain the injected value. Further tests show that we cannot reliably retrieve values of $\coo=200-300$, suggesting it is harder to detect C$^{18}$O compared to H$_2^{18}$O with our KPIC spectra. From examining the C$^{18}$O opacities, we find that this is may be due to the fact that many C$^{18}$O lines overlap in wavelength with $^{13}$CO lines, thereby masking the weaker signal from C$^{18}$O.

From the injection-recovery tests, we find that C/O and $\rm [C/H]$ are well-recovered, with only $\sim0.01$ deviations in C/O between the injected and recovered values, and $<0.02$ dex deviations in the $\rm [C/H]$ values. We attribute the slight offsets between the injected and retrieved values to random noise in the background trace. 

As mentioned in \S~\ref{sec:retrieval_setup}, the choice of rotational broadening kernel can have an impact on the retrieved \vsini. Even when using the direct integration method from \citet{Carvalho_2023}, we note that the retrieved \vsini can still be biased by the down-sampling factor for the line-by-line opacities. At the KPIC resolution of $R\sim35,000$ and for a \vsini of $5.0$ km/s (technically below our resolution limit of $\sim7.5$ km/s), we find that down-sampling the native $R=10^6$ opacities by a factor of 3 or less allows us to accurately recover the input \vsini of $5.0$ km/s.

\begin{figure*}[t!]
    \centering
\includegraphics[width=\linewidth]{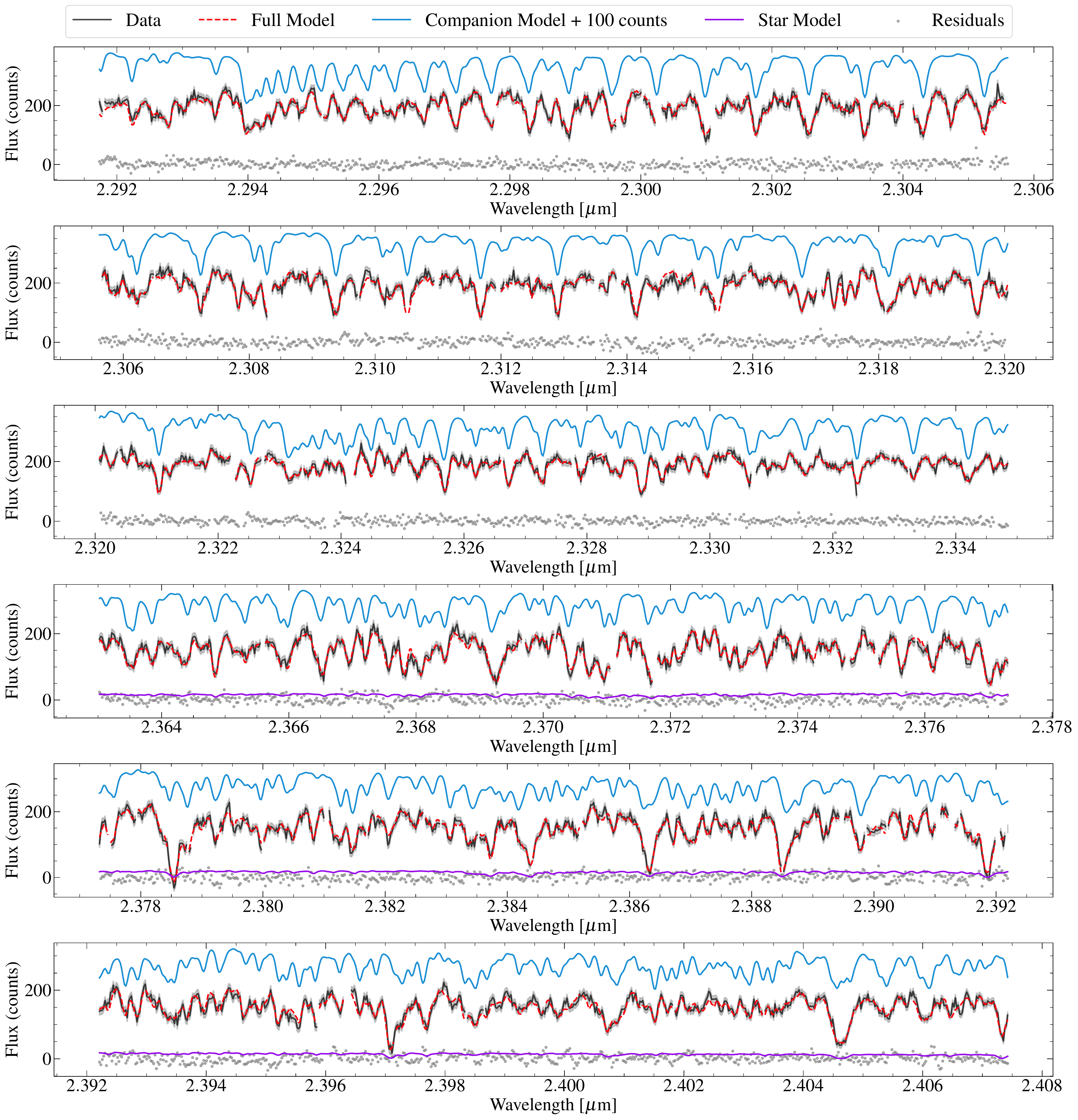}
    \caption{KPIC data (2023, fiber 2) for HIP~55507~B are plotted in black, with error bars in gray. We plot two out of three spectral orders ($2.29-2.41~\mu$m), and each order is broken into three panels. The full model ($FM_B$ in Eq.~\ref{eqn:comp}) is in red, and consists of the companion model ($M_c$) in blue, which has been RV shifted and broadened, the stellar spectra ($D_s$) in purple to model the speckle contribution, and the telluric and instrumental response ($T$). The fringing model is also incorporated in the full model. The residuals are shown as gray points. For clarity, an offset of +100 counts was added to the companion model. The speckle contribution is small, and consistent with zero for the first spectral order (top three panels), where we omitted the purple line.}
    \label{fig:kpic_model}
\end{figure*}

\begin{figure}[t!]
    \centering
\includegraphics[width=\linewidth]{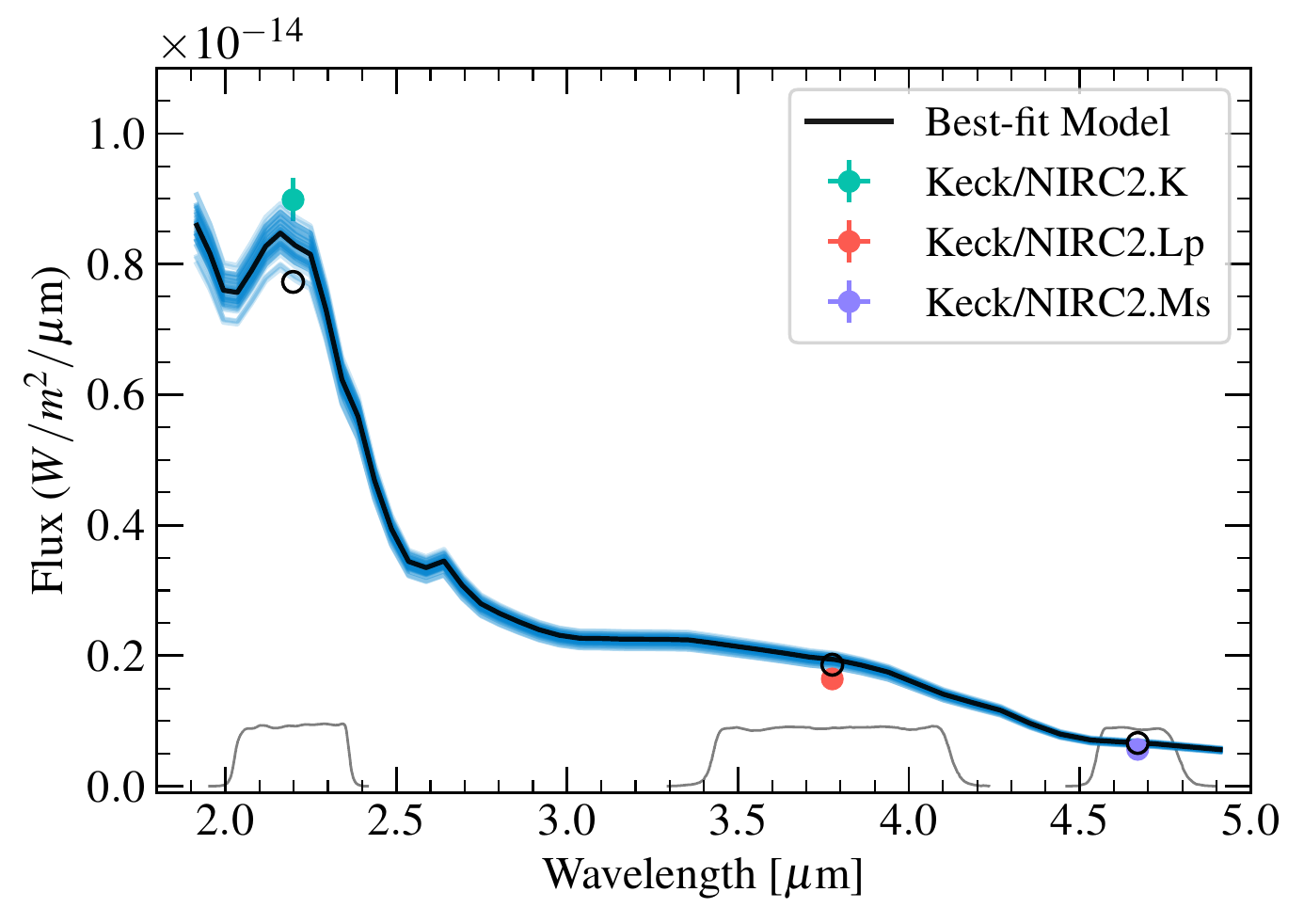}
    \caption{The photometry part of the retrieval for HIP~55507~B. The data are plotted in colored points, with filter transmission functions shown below. The best-fit model photometry points are shown in black open circles. The black curve is the best-fit spectra underlying the model photometry, while blue curves are random draws from the posterior.}
    \label{fig:phot_model}
\end{figure}

\begin{figure}
    \centering
\includegraphics[width=\linewidth]{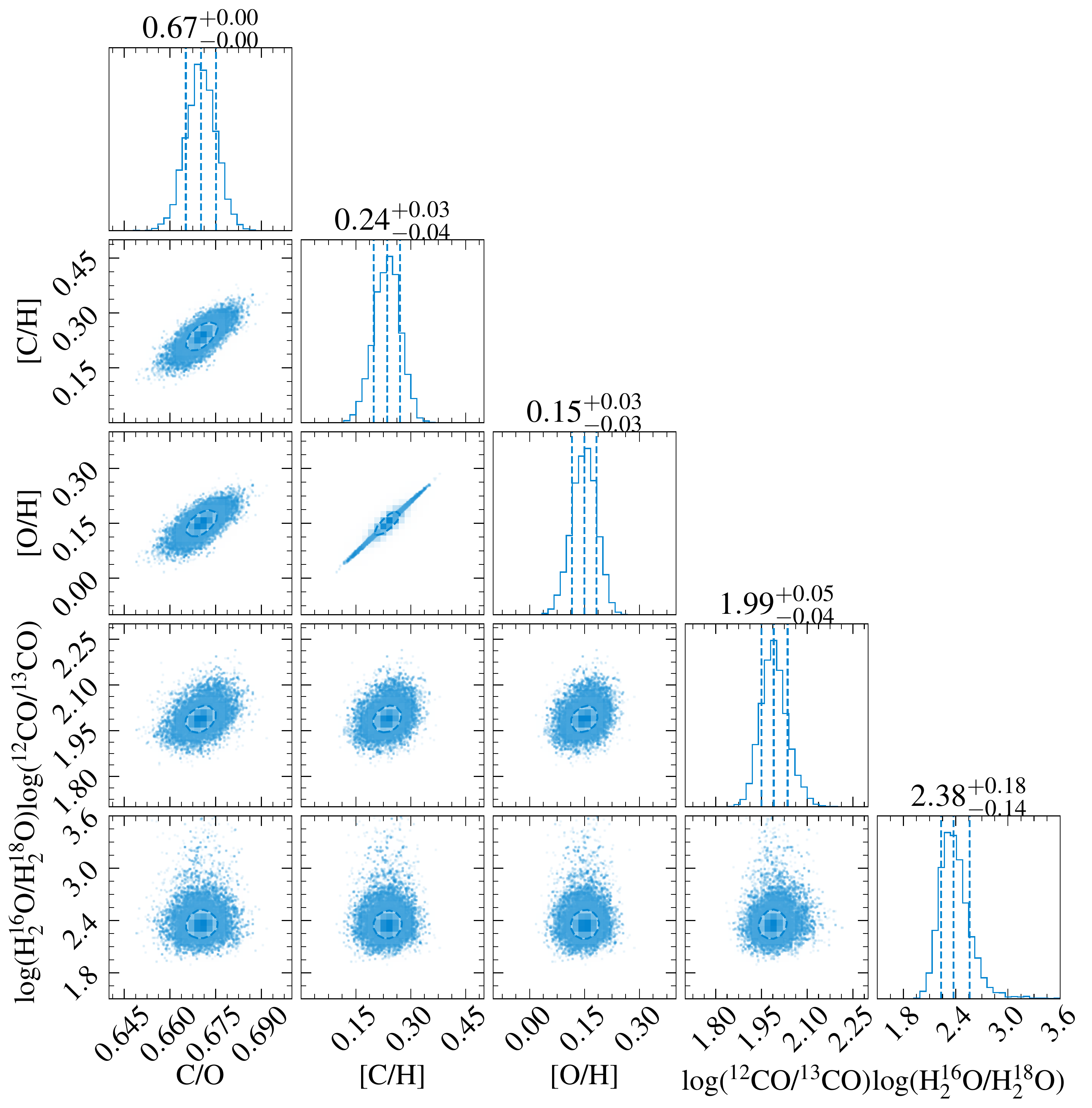}
    \caption{Posterior distributions for five key parameters from the baseline retrieval for HIP~55507~B. The titles on each histogram show the median and 68\% credible interval. These represent the statistical errors, and we account for systematic errors in our reported values in Table~\ref{table:spec_res}. Note that $\rm [O/H]$ is not fitted in the retrievals, but calculated from the C/O and $\rm [C/H]$ posteriors. The tight correlation between $\rm [C/H]$ and $\rm [O/H]$ indicates that KPIC high-resolution spectra can constrain relative abundances to much higher precision than absolute abundances, as found by previous high-resolution studies \citep{Xuan2022, Finnerty2023}.}
    \label{fig:corner_twodate}
\end{figure}

\section{Retrieval results}\label{sec:retrieval_results}

\subsection{HIP~55507~B}
We run three sets of retrievals for HIP~55507~B: two sets for the 2021 and 2023 KPIC datasets separately, and one set that combines the 2021 and 2023 KPIC data. By default, we include the $K$, $L^\prime$, and $M_s$ photometric data in the retrievals, but we also tested fits where we excluded the photometry. Retrieved parameters from each set of retrievals are presented in Table~\ref{table:spec_res}, and the baseline retrieval (2021+2023+photometry) is in bold. We plot the KPIC spectra and best-fit models in Fig.~\ref{fig:kpic_model}, while the photometry fit is shown in Fig.~\ref{fig:phot_model}. The joint posterior distributions of a few parameters from the baseline retrieval are shown in Fig.~\ref{fig:corner_twodate}.

\begin{deluxetable*}{cc|cccccc|c}[t!]\label{table:spec_res}
\tablecaption{Spectral Retrievals and Results}
\tabletypesize{\footnotesize}
\tablehead{
Data/Observing date & Isotopologues included & C/O & $\rm [C/H]$ & \co & \ho & Radius (\Rj) & $T_\textrm{eff}$ (K) & ln($B$)}
\startdata
\hline
2023/05/02 \\
KPIC + Phot. & $^{13}$CO, H$_2^{18}$O, C$^{18}$O & $0.68\pm0.01$ & $0.26\pm0.04$ & $106_{-11}^{+14}$ & $165_{-44}^{+68}$ & $1.32\pm0.02$ & $2367\pm20$ & 0 \\
KPIC + Phot. & H$_2^{18}$O, C$^{18}$O  & $0.68\pm0.01$ & $0.26\pm0.04$ & ... & $150_{-38}^{+87}$ & $1.32\pm0.02$ & $2355\pm20$ & -28.0 \\  
KPIC + Phot. & $^{13}$CO, C$^{18}$O & $0.68\pm0.01$ & $0.25\pm0.05$ & $103_{-9}^{+12}$ & ... & $1.32\pm0.02$ & $2372\pm20$ & -4.9 \\ 
KPIC + Phot. (Gray) & $^{13}$CO, H$_2^{18}$O, C$^{18}$O & $0.68\pm0.01$ & $0.25\pm0.04$ & $106_{-11}^{+13}$ & $163_{-42}^{+67}$ & $1.32\pm0.02$ & $2366\pm20$ & 2.5 \\  
KPIC + Phot. (Al$_2$O$_3$) & $^{13}$CO, H$_2^{18}$O, C$^{18}$O & $0.68\pm0.01$ & $0.25\pm0.05$ & $103_{-9}^{+13}$ & $167_{-46}^{+72}$ & $1.32\pm0.02$ & $2368\pm20$ & 2.3 \\  
KPIC & $^{13}$CO, H$_2^{18}$O, C$^{18}$O & $0.69\pm0.01$ & $0.00\pm0.12$ & $114_{-12}^{+15}$ & $172_{-43}^{+63}$ & $1.92\pm0.27$ & $2348\pm20$ & ... \\
\hline
\hline
2021/07/04 \\ 
KPIC + Phot. & $^{13}$CO, H$_2^{18}$O,  C$^{18}$O & $0.64\pm0.01$ & $0.14\pm0.05$ & $63_{-8}^{+13}$ & $>120~(3\sigma)$ & $1.35\pm0.02$ & $2343\pm25$ & 0  \\
KPIC + Phot. & H$_2^{18}$O, C$^{18}$O & $0.64\pm0.01$ & $0.12\pm0.05$ & ... & $>96~(3\sigma)$ & $1.35\pm0.02$ & $2351\pm25$ & -14.8 \\  
\hline
\hline
2021/07/04 + 2023/05/02 \\
\textbf{Adopted}: KPIC + Phot. & $^{13}$CO, H$_2^{18}$O, C$^{18}$O & $0.67\pm0.04$ & $0.24\pm0.13$ & $98_{-22}^{+28}$ & $240_{-80}^{+145}$ & $1.33\pm0.02$ & $2350\pm50$ & ... \\
\hline
\enddata
\tablecomments{A few atmospheric parameters and their central 68\% credible interval with equal probability above and below the median are listed. These values only account for statistical error. In the final row, we list the adopted values accounting for systematic errors from the retrieval. The rightmost column lists the log Bayes factor ln($B$) for each retrieval. We compute ln($B$) with respect to the baseline model for each dataset, i.e. the models with ln($B$)=0. Unless specified in parentheses, we use a clear model. `Gray' refers to the gray opacity cloud model, and `Al$_2$O$_3$' refers to the EddySed model with Al$_2$O$_3$ clouds.}
\end{deluxetable*}

\subsubsection{P-T profile and clouds}
The P-T profile from the baseline retrieval (clear model) is shown in Fig.~\ref{fig:ptprofile}. The cloudy models give almost identical P-T shapes as the clear model. We find that the lower atmosphere is fairly consistent with the self-consistent SPHINX models from \citet{Iyer2023}, but the upper atmosphere is hotter and more isothermal. This could be due to a trade-off between clouds and an isothermal P-T profile, which is seen in most retrieval studies \citep[e.g.][]{burningham_retrieval_2017, Xuan2022, BrownSevilla2023, Whiteford2023}. As demonstrated by \citet{Xuan2022}, narrow-band high-resolution spectra can be largely insensitive to clouds but still sensitive to gas-phase molecular abundances. To check if the isothermal upper P-T profile affects our results, we ran a retrieval for HIP~55507~B with the P-T profile fixed to the $\Teff=2400~K$, $\logg=5.25$ SPHINX profile in Fig.~\ref{fig:ptprofile}. The resulting posteriors from this retrieval are consistent with the those from our baseline retrieval within $1\sigma$, so we conclude that the isothermal upper atmosphere is not biasing our results.

For the 2023 KPIC dataset, we tested three clouds models: clear, gray opacity, and EddySed with Al$_2$O$_3$. To assess whether clouds are preferred by the data, we use the Bayesian evidence from each retrieval to calculate the Bayes factor $B$, which assesses the relative probability of an alternative model $M_2$ compared to $M_1$. Here, we take the clear model to be $M_1$. The data slightly prefer the gray opacity and EddySed models over the clear model, with ln($B$)=2.5 and ln($B$)=2.3, respectively, which correspond to $\sim2.6\sigma$ preferences for the cloudy models using the \citet{Trotta2008} scale. However, the cloud parameters are largely unconstrained, with a $3\sigma$ upper limit of $0.009~\rm{cm}^2/g$ for the gray opacity. The retrieved abundances are also identical between the cloudy and clear models (see Table~\ref{table:spec_res}), so we adopt the clear model as the baseline model. 

\begin{figure*}
    \centering
  \includegraphics[width=.45\linewidth]{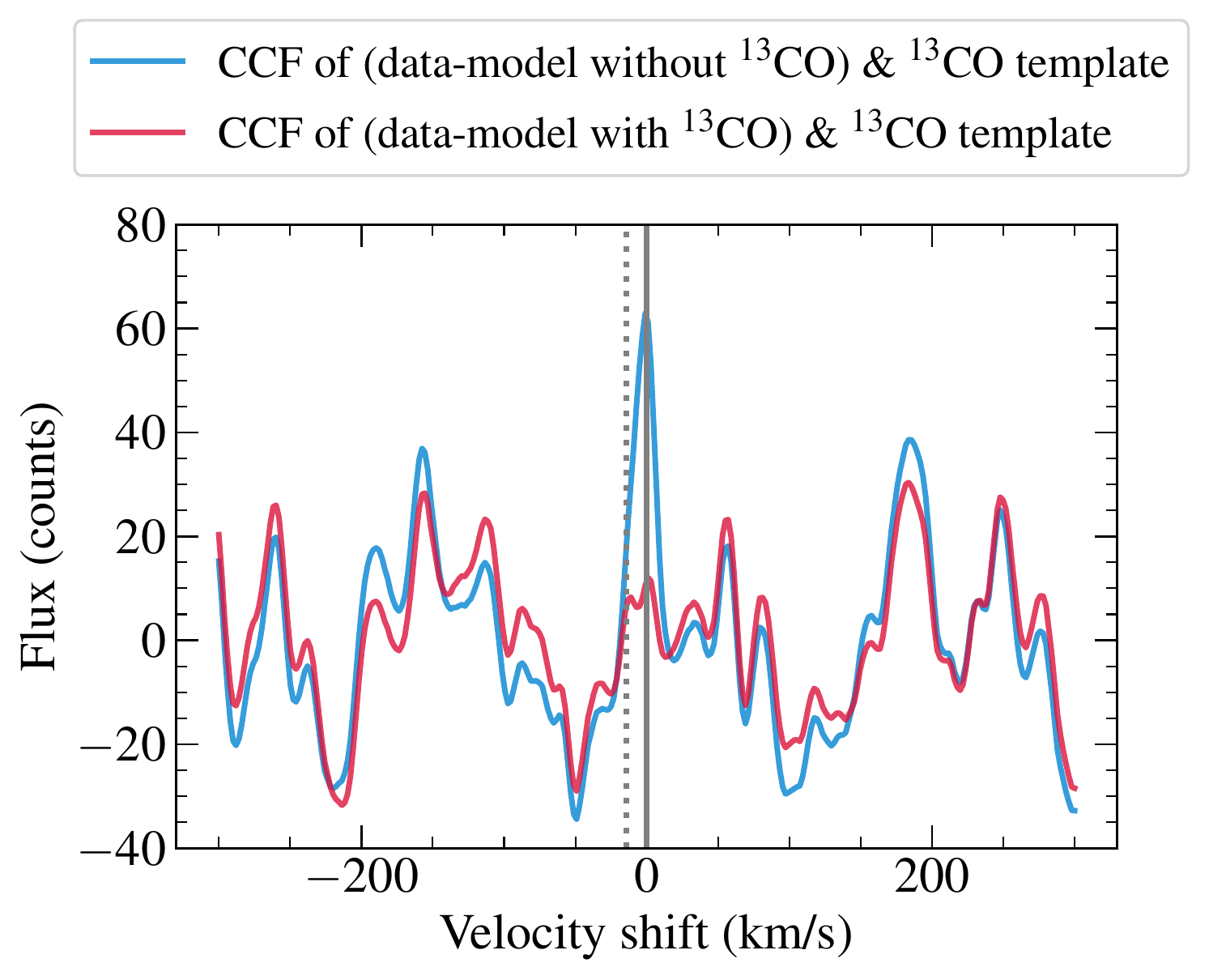}
    \centering
  \includegraphics[width=.45\linewidth]{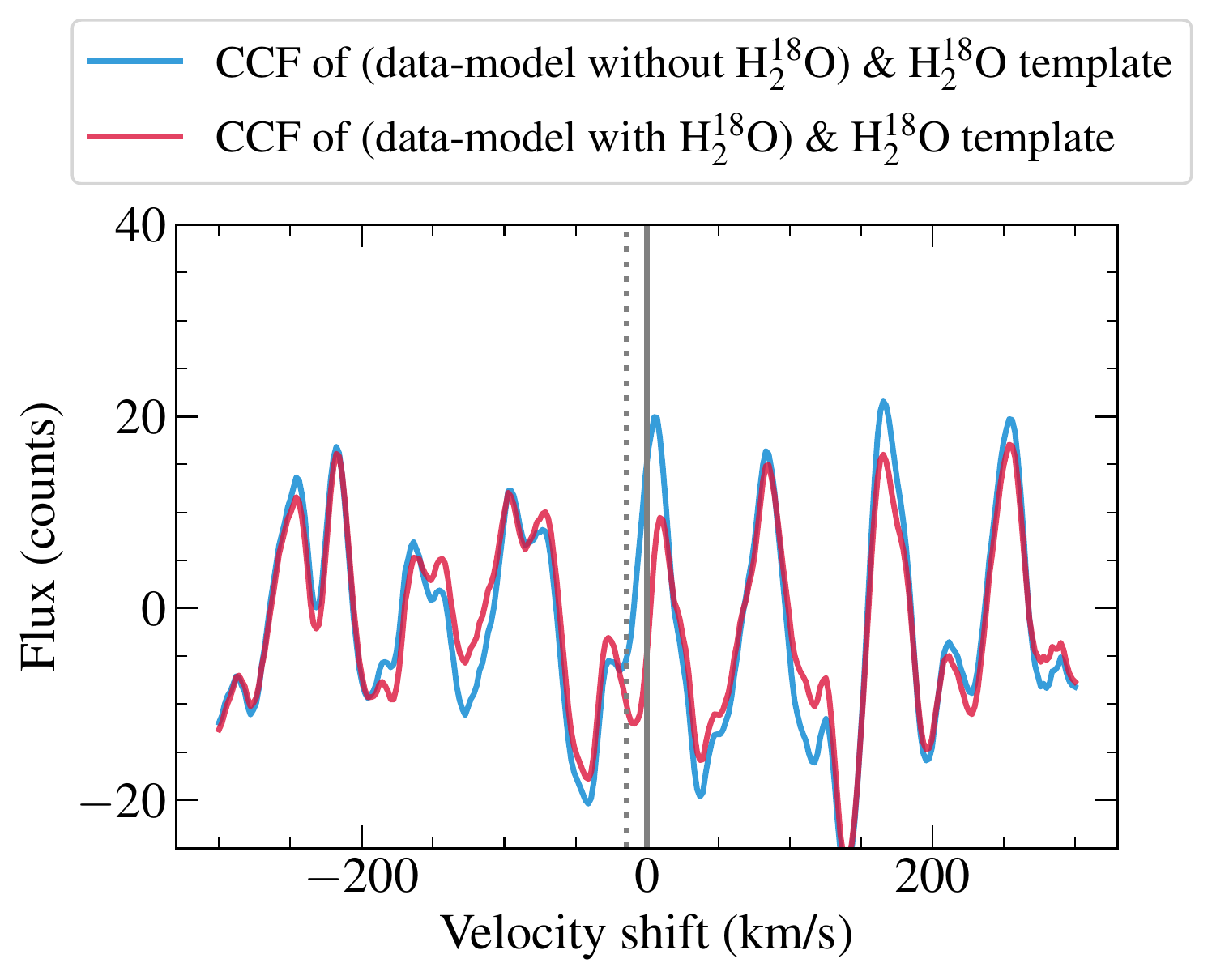}
    \caption{Left panel: The CCF between the $\rm{^{13}CO}$-only template and (data - best-fit model without $\rm{^{13}CO}$) in blue. The CCF between the $\rm{^{13}CO}$-only template and (data - best-fit model with $\rm{^{13}CO}$) is shown in red. The fact that the blue CCF shows a peak at the companion's rest frame (gray solid line) indicates a real $\rm{^{13}CO}$ detection. For comparison, the gray dotted line is the telluric rest frame. In the red CCF, we do not expect a peak since $\rm{^{13}CO}$ is fitted for in this model, so the residuals should be free of $\rm{^{13}CO}$. Right panel: same but for $\rm{H_2^{18}O}$. The $\rm{^{13}CO}$ signal is much clearer than the $\rm{H_2^{18}O}$ signal, which remains tentative at this stage.} \label{fig:iso_B_ccfs}
\end{figure*}

\subsubsection{Isotopologue abundances of $\rm{^{13}CO}$ and $\rm{H_2^{18}O}$}\label{sec:isoB}
In the KPIC dataset from 2023, we strongly detect $\rm{^{13}CO}$ and tentatively detect $\rm{H_2^{18}O}$ in the atmosphere of HIP~55507~B, with $\logco=2.03\pm0.05$ and $\logho=2.22\pm0.15$. The 2021 epoch gives $\logco=1.80\pm0.08$, which is 0.23 dex or $\approx2.4\sigma$ lower than the 2023 value. \ho~is unconstrained from the 2021 epoch due to its lower S/N. The discrepancy in \logco~between datasets is somewhat higher than the $0.05-0.1$ dex systematic bias we identify in our injection-recovery tests. In our reported values, which are based on a joint fit to both 2021 and 2023 datasets, we add a 0.1 dex systematic error in quadrature to the measurement errors. In summary, we report $\co=98_{-22}^{+28}$ and $\ho=240_{-80}^{+145}$ for HIP~55507~B. $\coo$~is unconstrained, with a formal $3\sigma$ lower limit of $\approx440$. However, as discussed in \S~\ref{sec:inj_rec}, we often cannot recover $\coo$ values of 200-300 from injection-recovery tests, so the true $\coo$ value could in fact be lower than 440 and consistent with our \ho~value. 

To assess whether the $^{13}$CO and H$_2^{18}$O isotopologues are needed to fit the data or whether we can adjust other parameters to  improve the fits, we perform two retrievals with one of these isotopologues removed in each. These constitute the `reduced models'. We then calculate the Bayes factor between these reduced models and the full model with all isotopologues included, which are tabulated in Table~\ref{table:spec_res}. The ln($B$) values correspond to a $7.8\sigma$ detection of $\rm{^{13}CO}$ and $3.7\sigma$ detection of $\rm{H_2^{18}O}$. 

We can obtain a complementary perspective on the robustness of these detections using the cross-correlation method, following the approach in \citet{zhang_12CO_2021} and \citet{Xuan2022}. The goal of this analysis is to assess whether the full models prefer $\rm{^{13}CO}$ and $\rm{H_2^{18}O}$ independent of the Bayes factor calculation from our retrievals. To compute the cross-correlation function (CCF), we follow the framework from \citet{Ruffio2019}, so the y-axis of our CCFs is the estimated flux level (in counts) of the isotopologue signal from a least-squares minimization.

First, we compute the CCF between a $\rm{^{13}CO}$-only model and the (data - model without $\rm{^{13}CO}$). The latter is equivalent to the residuals of the reduced model, and will contain residual $\rm{^{13}CO}$ lines if the data contains $\rm{^{13}CO}$. Then, we compute the CCF between the $\rm{^{13}CO}$-only model and the (data - model with $\rm{^{13}CO}$). This second CCF should not show a detection, as $\rm{^{13}CO}$ is already fitted for in the model with $\rm{^{13}CO}$ (i.e. the full model). We generate the isotopologue-only models by manually zeroing the opacities of all other line species except the isotopologue when computing the full model. The same process is repeated for $\rm{H_2^{18}O}$, whose CCFs are shown in the right panel of Fig.~\ref{fig:iso_B_ccfs}.

We find that the $\rm{^{13}CO}$ signal is cleaner compared to the $\rm{H_2^{18}O}$ signal, as the CCF for $\rm{H_2^{18}O}$ shows stronger residual structure in the wings, although there is a peak around 0 km/s (companion's rest frame) consistent with a real signal. However, because the remaining systematics are on a scale similar to that of the peak, we consider the $\rm{H_2^{18}O}$ detection to be tentative in these data. It is possible that this detection is produced by remnant fringing features that our fringe model did not perfectly remove and/or residual telluric features, which are especially strong in the $2.45-2.49$ wavelength region where the $\rm{H_2^{18}O}$ lines are strongest. A future upgrade to KPIC should greatly reduce the fringing from the dichroics, allowing us to re-visit $\rm{H_2^{18}O}$ in HIP~55507~B's atmosphere with more confidence.

\subsubsection{C/O and atmospheric metallicity}
We retrieve C/O=$0.68\pm0.01$, $\rm [C/H]=0.26\pm0.04$ for the 2023 epoch and C/O=$0.64\pm0.01$, $\rm [C/H]=0.14\pm0.05$ for the 2021 epoch. These values are broadly consistent, with a $\sim6\%$ difference in C/O and 0.12 dex difference in $\rm [C/H]$. Notably, the $\sim6\%$ difference in C/O between the two epochs is lower than the $\sim15-20\%$ error estimated by \citet{Xuan2022} and \citet{wang_Retrieving_2022} for benchmark BD companions, where it was assumed that the BD companions have the same compositions as their host stars. 

We adopt a systematic error of 0.04 in C/O and 0.12 dex in $\rm [C/H]$ for our baseline retrieval, and report $\rm [C/H]=0.24\pm0.13$, $\rm [O/H]=0.15\pm0.13$, and C/O=$0.67\pm0.04$ for HIP~55507~B. The primary star HIP~55507~A has [Fe/H] = $-0.02\pm0.09$ from Keck/HIRES spectra in the optical (Table~\ref{tab:prop}), consistent with a solar metallicity. If we assume $\rm [Fe/H]=[C/H]$ for the primary star, this implies the $\rm [C/H]$ between HIP~55507~A and B are consistent to within $1.6\sigma$.

\subsubsection{Effective temperature, luminosity, and radius}\label{sec:teff_rad}
The addition of the photometry data in $K$, $L^{\prime}$, and $M_s$ bands helps constrain the bulk parameters of HIP~55507~B. For example, when we omitted the photometry for the May 2023 retrieval, the retrieved radius is $R=1.92\pm0.27~\Rj$, while retrievals with the the photometry yield $R=1.32\pm0.02~\Rj$ (Table~\ref{table:spec_res}). We integrate model spectra with parameters drawn from the posteriors of our baseline retrieval to compute $\lbollsun=-3.29\pm0.02$ and $\Teff=2350\pm50$~K. While the statistical uncertainties on these parameters are small, we note that the model uncertainties are likely larger since the flux information is derived from three photometric points covering a small portion of the \lbol~budget.

Given measurements of the dynamical mass and \lbol, we can compare the spectrally-inferred radius and \Teff to the predictions from evolutionary models. To do so, we interpolate the BHAC15, AMES-COND, and AMES-Dusty models \citep{Baraffe2015, Allard2001} with Gaussian distributions of $m=88.0\pm3.4~\Mj$ and $\lbollsun=-3.29\pm0.02$. We find that the evolutionary models favor $R\approx1.08\pm0.02~\Rj$ and $\Teff=2530\pm80~K$, i.e. a smaller radius and larger \Teff than the spectral retrievals. A similar radius-\Teff~degeneracy has been noted by several retrieval studies, although the discrepancy is usually in the opposite direction for colder brown dwarfs, with retrievals finding a smaller radius than evolutionary models \citep[e.g.][]{Lueber2022, Gonzales2022, Hood2023}. \citet{Sanghi2023} compared radii inferred by evolutionary models and atmospheric models for a large sample of brown dwarfs and found significant discrepancies in \Teff and radius for late-M/early-L spectral types in both directions, which highlights ongoing challenges in measuring bulk properties from substellar atmospheric models and retrievals (see also \citealt{Dupuy2010} for late-M dwarfs).

\subsubsection{Projected rotation rate}
We measure a relatively slow $\vsini=5.4\pm0.2~$km/s for HIP~55507~B, which is below the KPIC resolution limit of $\sim7.5~$km/s at $R\sim35,000$. The high S/N of the data allow us to tightly constrain \vsini~values below the resolution limit, as we demonstrated using injection-recovery tests (Table~\ref{table:inj_rec}).

\subsection{HIP~55507~A}\label{sec:iso_A}
Using the higher S/N spectra from 2023, we carry out retrievals with the sole goal of measuring isotopic abundances in HIP~55507~A. The best-fit model and stellar spectra are shown in Appendix~\ref{app:A_retrieval_model}. We are able to measure \co~and \coo~in the photosphere of the primary star. We add the same 0.1 dex systematic error as we did for HIP~55507~B, resulting in $\co=79_{-16}^{+21}$ and $\coo=288_{-70}^{+125}$. 

In these HIP~55507~A retrievals, we assumed a fixed P-T profile ($\Teff=4300~K$, $\logg=4.5$). To assess the impact of this assumption on the resulting isotopic abundances, we repeated the fits with $\Teff=4200~K$ and $\Teff=4400~K$ P-T profiles (same \logg). We find that varying the P-T profile has little effect on the results; the posteriors shift by $<1\sigma$.

\subsection{Relative RVs between HIP~55507~AB}\label{sec:host_rvs}
By combining the retrievals for HIP~55507~A and B, we compute the relative RV between the stars at the time of observation. The RV values corrected for barycentric motion are provided in Appendix~\ref{app:relrv_star}. In both the 2021 and 2023 epochs, the stellar and companion RVs measured from fiber 3 are higher compared to those measured from fiber 2. This may be due to different RV zeropoints for each fiber. To compute the relative RV, we subtract the stellar RV from the companion RV for each fiber separately, and then take the average. For the 2023 epoch, the relative RV is consistent at the $<0.05$ km/s level between fibers. We conservatively adopt 0.1 km/s as the uncertainty. For the 2021 epoch, we adopt a larger uncertainty of 0.2 km/s as the relative RV values disagree by $\sim0.15$~km/s between fibers. 

\begin{figure}
    \centering
\includegraphics[width=\linewidth]{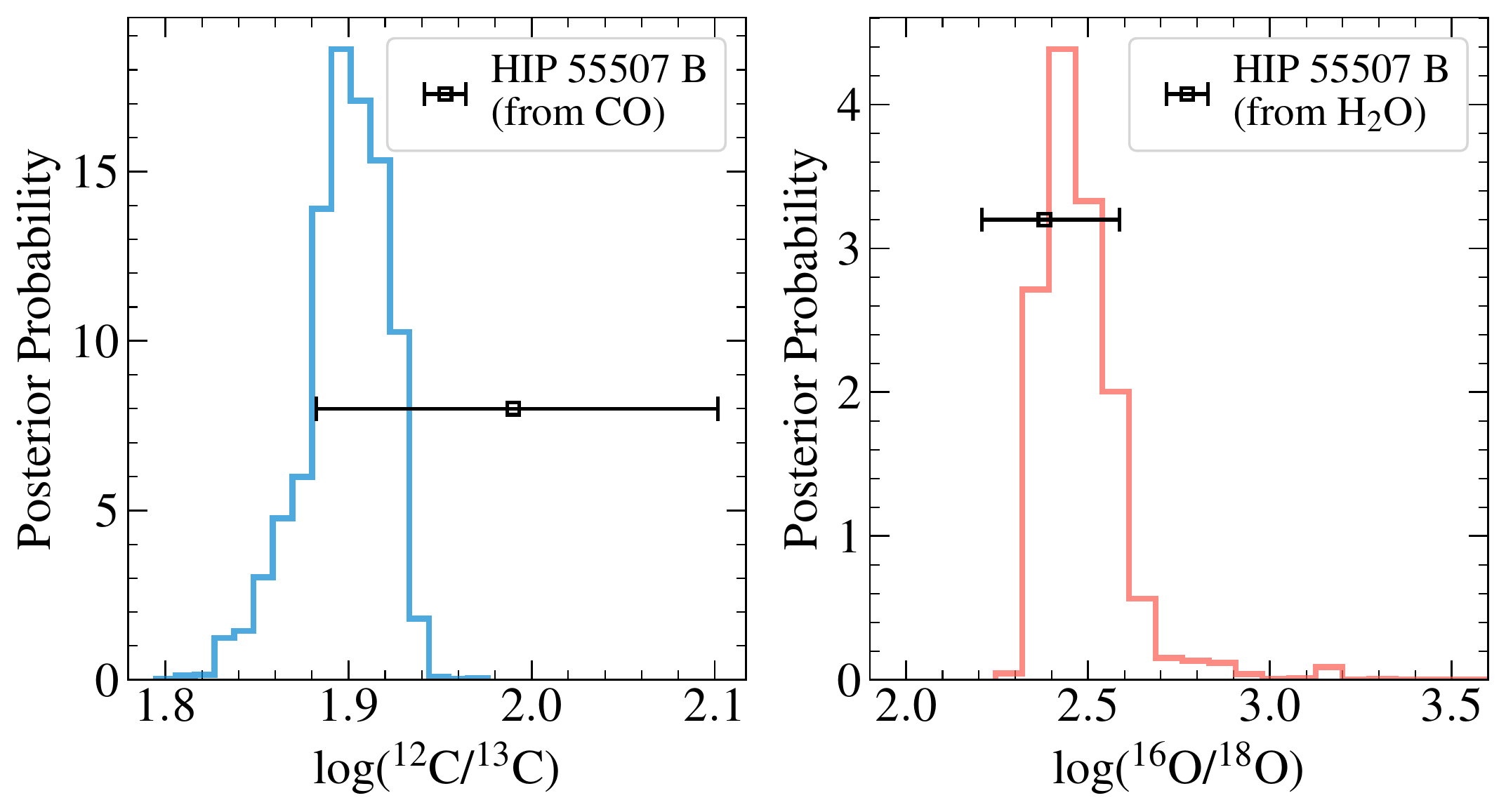}
    \caption{The posteriors for \logco~and \logcoo~measured from KPIC spectra of HIP~55507~A. We overplot the corresponding isotopic values measured for HIP~55507~B as error bars. Note that the \oratio~constraint for HIP~55507~B is from \logho.}
    \label{fig:iso_A_corner}
\end{figure}

\section{Discussion and Conclusions}\label{sec:conclude}

\subsection{Relative radial velocities and dynamical mass}
In this paper, we demonstrate the value of relative RV data in orbit fits, which are uniquely enabled by direct measurements of the companion RV \citep{Ruffio2019, Ruffio2023}. When the primary star has sufficient absorption lines in the $K$ band, as is the case for HIP~55507~A, we can directly measure the relative RV between the primary star and companion at the same epoch with Keck/KPIC. This measurement is powerful since it is insensitive to potential systematics from zeropoint offsets between different instruments used to acquire the spectra. By including two epochs of KPIC relative RVs in our orbit fit for HIP~55507~AB, we find a $60\%$ improvement in the dynamical mass measurement for HIP~55507~B (${88.0}_{-3.2}^{+3.4}~\Mj$). Dynamical masses of low-mass companions are key measurements that allow us to test evolutionary and atmospheric models. Future work should explore whether relative RVs can also improve constraints on companion mass in systems with longer orbital separations (and therefore less orbital coverage), as many directly imaged companions have orbital separations $\gtrsim100$ AU.

We note that our orbit solution for HIP~55507~AB also represents a significant improvement compared to those presented in \citet{Feng2022}, who did not use relative astrometry data from imaging and only had access to the first 6.3 years of HIRES RVs. Their derived orbital period of $14.0\pm1.4$ years and companion mass of $5.0\pm0.6~\Mj$ for HIP~55507~B is significantly discrepant from our results. Furthermore, the observed luminosity of the companion from our NIRC2 data is consistent with a low-mass star and not a $5.0~\Mj$ planet (see Fig.~\ref{fig:isochrone}). In addition, we cannot re-produce their results even if we used the same data as \citet{Feng2022}, namely the first 6.3 years of HIRES RVs and Gaia-Hipparcos absolute astrometry. In this case, our fits result in unbounded posteriors for mass and other orbital parameters (e.g. with a $1\sigma$ interval for semi-major axis from 13-840 AU). We conclude that the choice of prior ranges in the orbit fit may be biasing the \citet{Feng2022} results for HIP~55507~B.

\subsection{Isotopologue ratios}
Isotopologue ratios are thought to have implications for the formation pathway of planets and substellar companions, but our knowledge of how carbon and oxygen isotopic ratios relate to formation is still limited. We can benchmark the value of isotopologue measurements by using higher-mass brown dwarf and stellar companions which form via gravitational instability from a protostellar disk or molecular cloud. Because these systems are dominated by gas accretion, they should exhibit the same isotopic ratios between the primary and secondary components, which we can test observationally. To our knowledge, for main-sequence stellar binaries, this test has only been demonstrated once with a M dwarf binary system \citep{Crossfield2019}.

Using high-resolution spectra from Keck/KPIC ($R\sim35,000$), we perform atmospheric retrievals for the M7.5 companion HIP~55507~B ($\Teff\sim2400~K$) and its K6V primary star ($\Teff\sim4300~K$). For HIP~55507~B, we retrieve $\rm [C/H]=0.24\pm0.13$ and $\rm C/O=0.67\pm0.04$. As shown in Fig.~\ref{fig:iso_A_corner}, our measured $\co=98_{-22}^{+28}$ for HIP~55507~B is consistent to within $1\sigma$ with our measured $\co=79_{-16}^{+21}$ for the primary star under the assumption that HIP~55507~A and B share the same C/O and $\rm [C/H]$. Furthermore, the $\oratio~$measured from H$_2$O in HIP~55507~B is $240_{-80}^{+145}$, consistent with $\oratio=288_{-70}^{+125}$ measured from CO in HIP~55507~A. The agreement between \cratio~and \oratio~in the HIP~55507~AB system represents is a rare test of chemical homogeneity for stellar binaries using isotopic ratios.

We note that our value of $\ho=240_{-80}^{+145}$ for HIP~55507~B is lower by a factor of $\sim2$ compared to the solar value of $525\pm21$ \citep{Lyons2018}. A large difference between the \oratio~in HIP~55507~AB and the Sun is not unexpected, as measurements of \oratio~ in molecular clouds at the solar galactocentric distance  show a factor of $\sim3$ scatter \citep{Nittler2012}. Furthermore, studies of solar twins also reveal a wide range in \oratio, with values as low as $50-100$ \citep{Coria2023}. Given the tentative nature of the H$_2^{18}$O detection in HIP~55507~B however, follow-up observations would be needed to confirm this measurement. In addition, our KPIC spectra for HIP~55507~B does not have sufficient S/N to constrain \coo~even if \coo=\ho~(see \S~\ref{sec:inj_rec}). Measuring \coo~in HIP~55507~B consistent with \coo~in the primary star would be a valuable test of our retrieval method and another piece of evidence supporting isotopic homogeneity between HIP~55507~A and B.

With a case study of the M7.5 companion HIP~55507~B using Keck/KPIC spectroscopy, we demonstrate the ability to measure carbon and oxygen elemental and isotopic abundances for late-M spectral types. In addition, we use KPIC to measure \cratio~and \oratio~for its K6V primary star and confirm that the companion and primary share the same isotopic abundances. While we made simplifications in our analysis of HIP~55507~A, future work with more extensive wavelength coverage (e.g. $H$ + $K$ bands) could explore more sophisticated retrievals for late-K and early-M dwarf stars. Finally, the projected separation and flux ratio between HIP~55507~A and B are comparable to systems with young ($\sim1-50$ Myr) substellar companions of similar spectral types as HIP~55507~B ($\Teff\sim2000-2800~K$), which opens the door to a systematically measuring the elemental and isotopic abundances of these companions with KPIC.

\acknowledgements
J.X. thanks Paul Mollière for help with generating new opacities in \texttt{petitRADTRANS} and using \texttt{easyCHEM}. J.X. is supported by the NASA Future Investigators in NASA Earth and Space Science and Technology (FINESST) award \#80NSSC23K1434. J.X. also acknowledges support from the Keck Visiting Scholars Program (KVSP) to commission KPIC Phase II capabilities. D.E. is supported by the NASA FINESST award \#80NSSC19K1423. D.E. also acknowledges support from the Keck Visiting Scholars Program (KVSP) to install the Phase II upgrades.
Funding for KPIC has been provided by the California Institute of Technology, the Jet Propulsion Laboratory, the Heising-Simons Foundation (grants \#2015-129, \#2017-318, \#2019-1312, \#2023-4598), the Simons Foundation, and the NSF under grant AST-1611623.
The computations presented here were conducted in the Resnick High Performance Center, a facility supported by Resnick Sustainability Institute at the California Institute of Technology.
W. M. Keck Observatory access was supported by Northwestern University and the Center for Interdisciplinary Exploration and Research in Astrophysics (CIERA).
This research has made use of the Keck Observatory Archive (KOA), which is operated by the W. M. Keck Observatory and the NASA Exoplanet Science Institute (NExScI), under contract with the National Aeronautics and Space Administration.
The data presented herein were obtained at the W. M. Keck Observatory, which is operated as a scientific partnership among the California Institute of Technology, the University of California and the National Aeronautics and Space Administration. The Observatory was made possible by the generous financial support of the W. M. Keck Foundation. The authors wish to recognize and acknowledge the very significant cultural role and reverence that the summit of Mauna Kea has always had within the indigenous Hawaiian community.  We are most fortunate to have the opportunity to conduct observations from this mountain.

\facilities{Keck (KPIC)}
\software{\texttt{petitRADTRANS}~\citep{molliere_petitRADTRANS_2019}, \texttt{dynesty}~\citep{speagle_DYNESTY_2020}, \texttt{species}~\citep{Stolker2020}}

\clearpage

\appendix

\onecolumngrid

\FloatBarrier
\section{HIRES RVs for HIP~55507~A}\label{app:hires_rvs}

\restartappendixnumbering
\begin{deluxetable*}{ccc}
    \tablecaption{HIRES RV Measurements for HIP~55507~A}
    \label{tab:hires_rvs}
    \tablehead{\colhead{Epoch [JD]} & \colhead{RV [m/s]} & \colhead{RV error [m/s]}}
    \startdata
    2454928.956275 & -85.2891389311628 & 1.30891144275665 \\
    2455164.14972 & -68.7729316154637 & 1.47898721694946 \\
    2455188.145761 & -66.6667559203668 & 1.43427407741547 \\
    2455191.15537 & -75.0535222054357 & 1.24979019165039 \\
    2455232.042467 & -62.6764906469445 & 1.54032349586487 \\
    2455255.95588 & -72.1607865157624 & 1.34993493556976 \\
    2455260.916671 & -54.8581642759654 & 1.45500349998474 \\
    2455285.96214 & -59.8329068133204 & 1.38776302337646 \\
    2455342.836003 & -53.2058373492368 & 1.31268215179443 \\
    2455343.8461 & -57.5779338603028 & 1.33945858478546 \\
    2455344.865259 & -48.9945473322775 & 1.58772456645966 \\
    2455373.79232 & -62.9124972583049 & 1.31496453285217 \\
    2455376.787806 & -63.2145698783239 & 1.29381465911865 \\
    2455395.813169 & -50.7233486928718 & 1.46614670753479 \\
    2455399.805149 & -54.886460583263 & 1.60360312461853 \\
    2455557.052607 & -52.3534558265392 & 1.37891674041748 \\
    2455614.102829 & -43.3644036486916 & 1.48802018165588 \\
    2455635.058738 & -34.5544936031586 & 1.8173508644104 \\
    2455663.95977 & -44.8672651793666 & 1.25698328018188 \\
    2455668.877009 & -40.9864458731045 & 1.30355906486511 \\
    2455670.931216 & -42.3292736833009 & 1.37228858470917 \\
    2456320.142426 & 16.0774879418836 & 1.54433631896973 \\
    2456327.067151 & 7.77230437090594 & 1.60971903800964 \\
    2456450.81181 & 0.541037387102907 & 1.27542078495026 \\
    2457201.75372 & 49.0106801572104 & 1.33652639389038 \\
    2459373.76488 & 159.649231940397 & 1.27313685417175 \\
    2459541.113394 & 168.07387611986 & 1.4402596950531 \\
    2459546.067624 & 180.157805514196 & 1.30128860473633 \\
    2459592.990193 & 189.869930526855 & 1.37484121322632 \\
    2460094.770983 & 199.762668142396 & 1.5842090845108 \\
    2460104.749802 & 200.888299897594 & 1.35568201541901 \\
    \enddata
\end{deluxetable*}

\onecolumngrid
\clearpage

\FloatBarrier
\section{NIRC2 Imaging and Orbit fits for HIP~55507~AB}\label{app:nirc2_orbit}
\begin{figure*}[h]
    \centering
\includegraphics[width=0.75\linewidth]{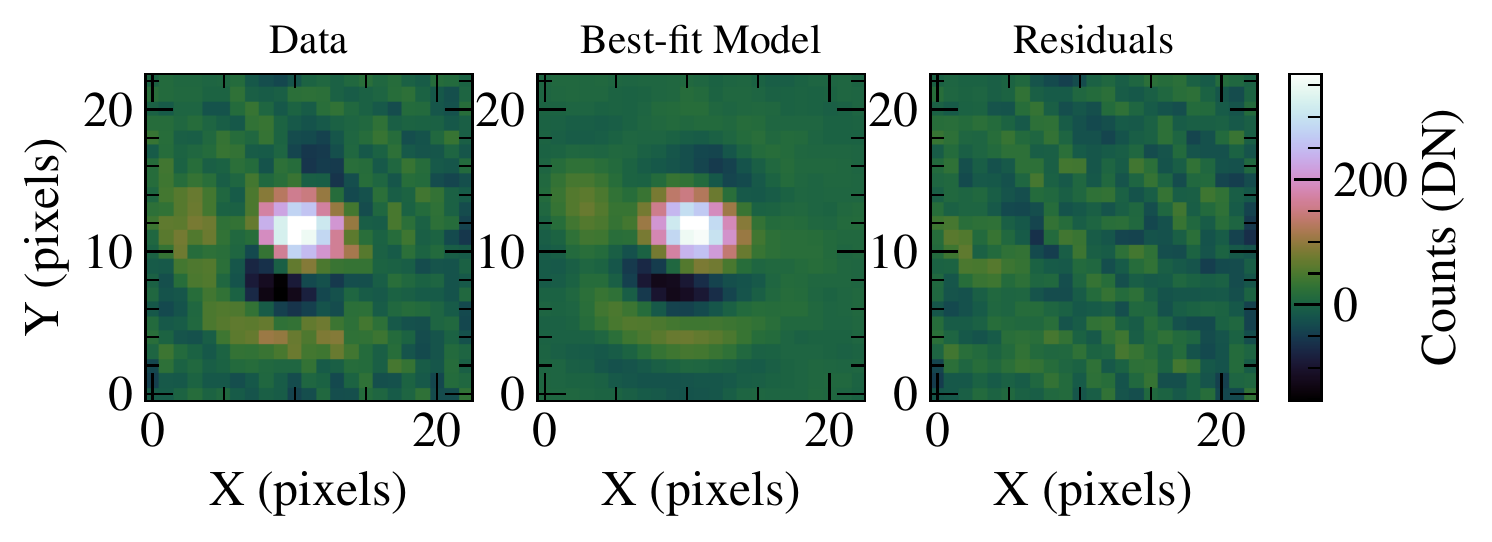}
    \caption{An example \texttt{pyklip} fit for the NIRC2 imaging data from UT 2022 June 9 with the $K$ filter. The companion PSF after ADI is shown on the left panel, the forward model in the middle panel, and residuals in the right panel.}
    \label{fig:Kband_nirc2_pyklip}
\end{figure*}

\begin{figure*}
\centering
\begin{subfigure}
  \centering
  \includegraphics[width=.4\linewidth]{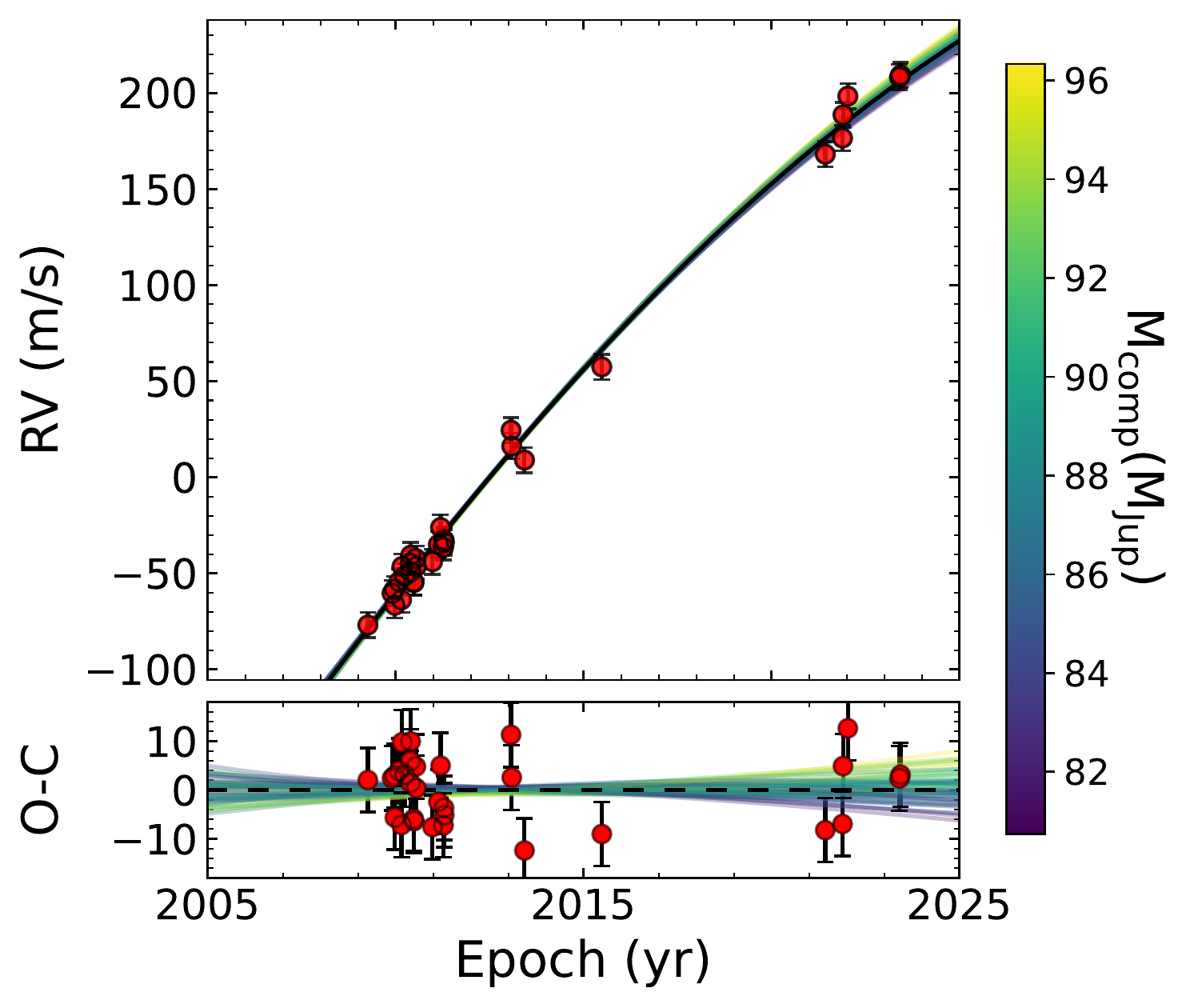}
\end{subfigure}
\begin{subfigure}
  \centering
  \includegraphics[width=.43\linewidth]{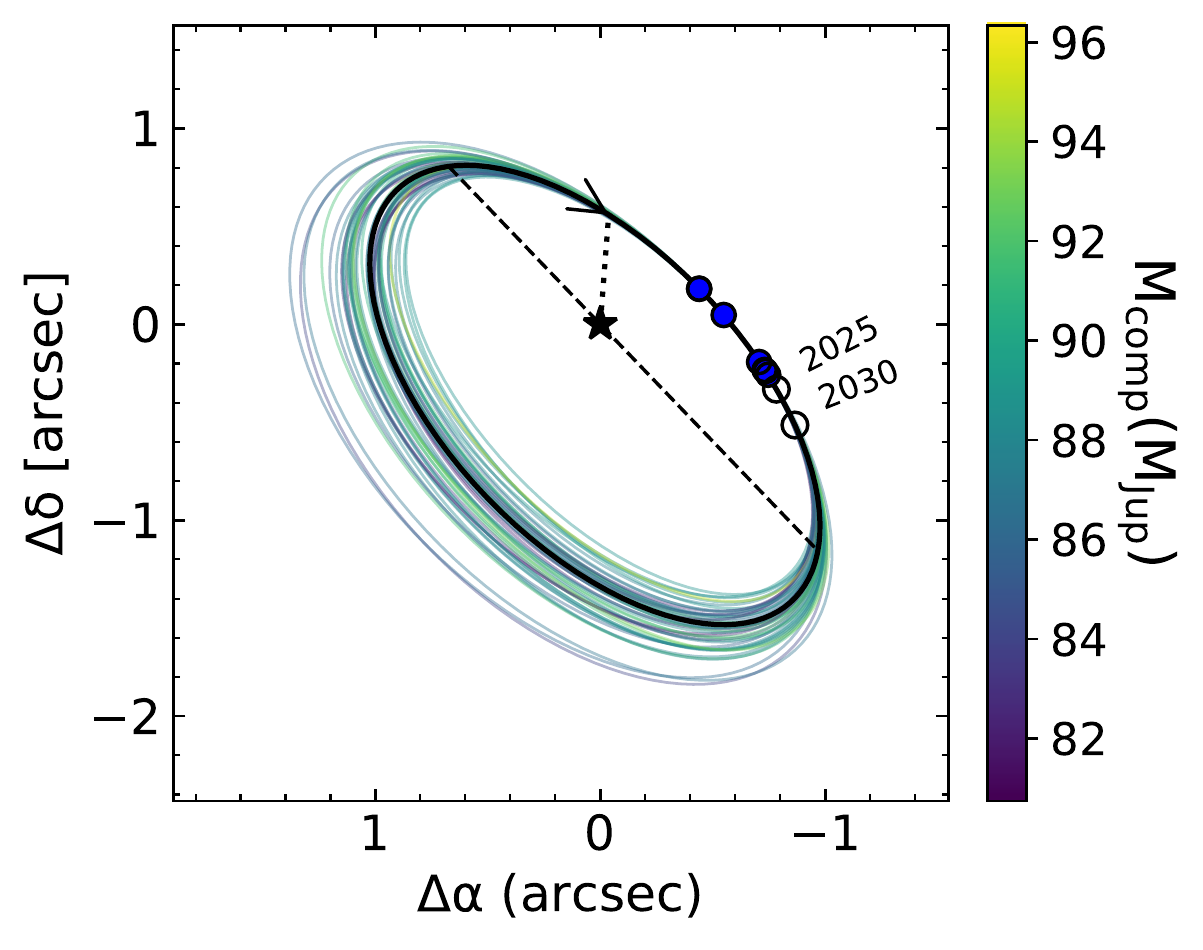}
\end{subfigure}
\begin{subfigure}
  \centering
  \includegraphics[width=.4\linewidth]{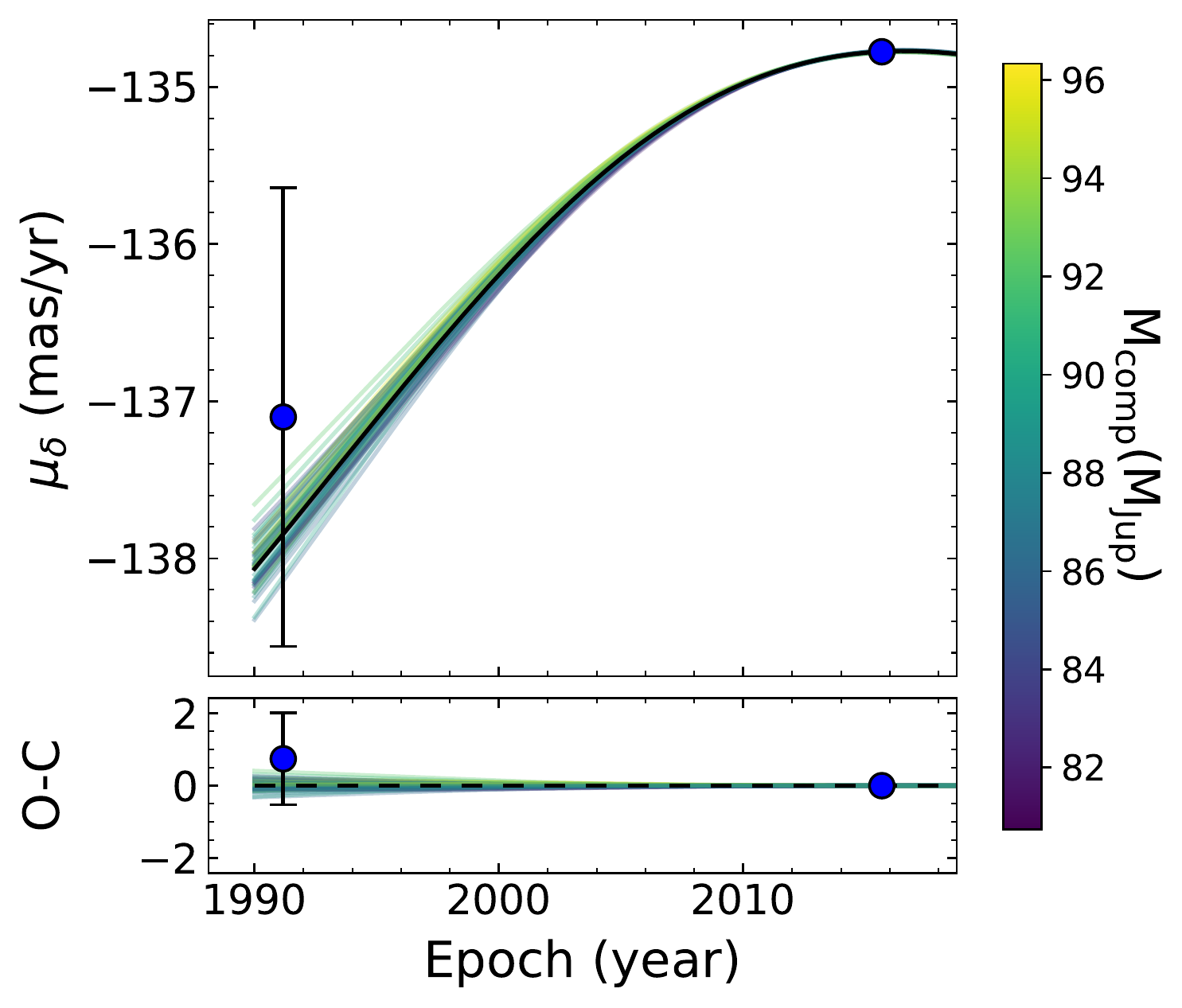}
\end{subfigure}
\begin{subfigure}
  \centering
  \includegraphics[width=.43\linewidth]{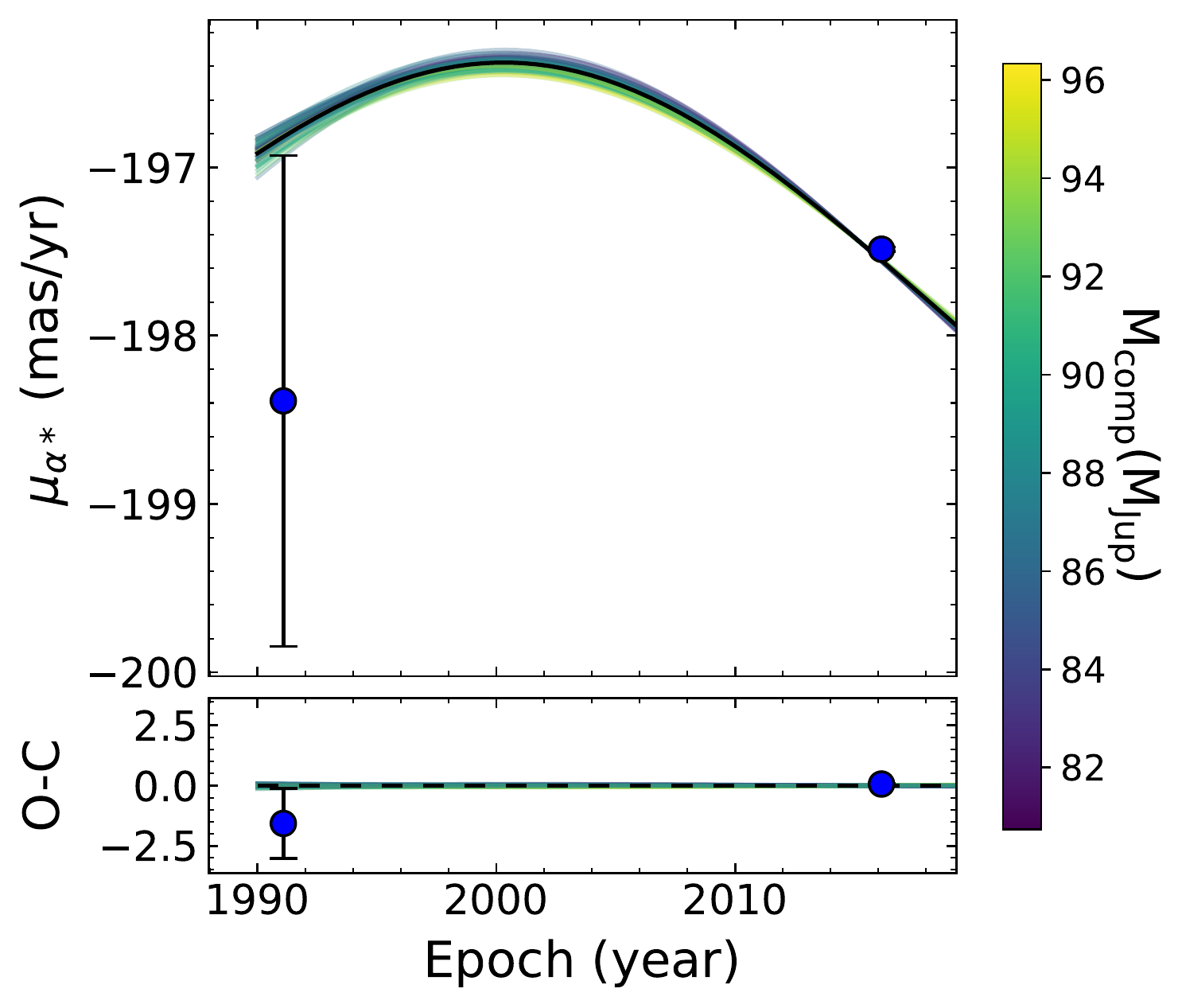}
\end{subfigure}
\caption{Results from the orbit fit using host star radial velocity (top left), relative astrometry (top right), absolute astrometry from Gaia and Hipparcos (bottom panels), and relative RVs from KPIC (shown in Fig.~\ref{fig:relrv}). The orbit fit is performed with the \texttt{orvara} package \citep{brandt_orvara_2021}. The random draws from the posterior are color-coded by the companion mass.}\label{fig:orbit}
\end{figure*}

\FloatBarrier
\section{KPIC RVs for HIP~55507~AB}\label{app:relrv_star}
\restartappendixnumbering

\begin{deluxetable*}{ccccc}[h]
    \tabletypesize{\footnotesize}
    \tablecaption{Radial Velocity Measurements for HIP~55507~A and B from KPIC. We have applied the barycentric correction to the individual RVs for A and B, so their reference is the solar system barycenter. The relative RV is defined as RV$_{\rm B}$ - RV$_{\rm A}$. For the relative RV values, we inflated the errors to account for systematics between fibers.}\label{tab:relrv_star}    
    \tablehead{\colhead{UT Date} & \colhead{Object} & \colhead{BJD-2400,000} & \colhead{SF2 RV (km/s)} & \colhead{SF3 RV} (km/s)}
    \startdata
    \hline
       2021 July 4 & HIP~55507~A & 59399.73 & $-5.48\pm0.03$ & $-5.38\pm0.03$ \\
       2021 July 4 & HIP~55507~B & 59399.73 & $-7.33\pm0.06$ & $-7.08\pm0.07$ \\
        Relative RV = $-1.78\pm0.20$ km/s \\
        \hline
        2023 May 2 & HIP~55507~A & 60066.78 & $-5.47\pm0.02$ & $-5.15\pm0.02$ \\
        2023 May 2 & HIP~55507~B & 60066.78 & $-7.31\pm0.05$ & $-7.03\pm0.06$ \\
        Relative RV = $-1.86\pm0.10$ km/s \\
    \enddata
\end{deluxetable*}

\FloatBarrier
\section{Fringing model for KPIC}\label{app:fringeA}
We identified three sources of fringing in KPIC data, (two dichroics in KPIC, and optics in the NIRSPEC entrance window \cite{Finnerty2022}. One of the two dichroics in particular causes the fringing signal to change when we switch from observing the primary star to observing an off-axis companion, as this dichroic is directly downstream of our fiber injection unit tip-tilt mirror which steers the light of either the star or companion in the fiber. When the tip-tilt mirror switches from the on-axis star to the off-axis companion, the angle of incidence of light into the dichroic changes, which causes the fringing signal to change. The change in modulation, $t$, as light passes through a transmissive optic is described by the well known formula:
\begin{equation}
    t = \left[1+F\sin^2\left(\frac{2\pi nl\cos(\theta)}{\lambda}\right) \right]^{-1}.
\end{equation}
Here, $F = 4R/(1-R)^2$ where $R$ is the reflectivity of the material, $n$ is the index of refraction of the material which depends on temperature and wavelength, $l$ is the thickness of the material, $\theta$ is the angle of incidence into the material, and $\lambda$ is the wavelength of light. All KPIC observations to date have spectra affected by three fringing modulation terms multiplied in series, but two of them are expected to be relatively static when going from star observations to companion observations. 

We fit the fringing signal in these spectra with a simplified approximation where the companion observations experience an additional modulation term as compared to the star observations. We also simplify the above equation to:

\begin{equation}
    t' = \left[1+F\sin^2\left(\frac{2\pi n(T_{d},\lambda) \times d}{\lambda}\right) \right]^{-1}.
\end{equation}
We multiply $t'$ onto the spectral response (i.e. $T$ in Eq.~\ref{eqn:comp} and Eq.~\ref{eqn:star}) to match the fringing in the observed spectra for both HIP~55507~A and B. We fit for three parameters per spectral order: an optical path length ($d$) term that combines both the thickness of the glass and the angle of incidence, the fractional amplitude of the ghost from the dichroics ($F$), and the temperature of the dichroics ($T_d$) that governs the index of refraction. To model the CaF$_2$ dichroic we used the Sellmeier coefficients reported by \cite{Leviton2008_Caf2} to determine how the index of refraction changes with wavelength and temperature. Each science fiber is treated separately, as the fringing is different in each due to different angles of incidence. 

\FloatBarrier
\section{Fitting HIP~55507~A spectra with PHOENIX-ACES models}\label{app:phoenix_A}
We fit the HIP~55507~A spectra using PHOENIX-ACES models \citep{husser_new_2013} to measure RV, \Teff, and \logg. Our grid of PHOENIX-ACES models assumes solar metallicity and has 100~$K$ spacing in \Teff and 0.5 dex spacing in \logg. The forward model and fringing model used for this fit are described in \S~\ref{sec:fm_kpic}.

From these fits, we obtain a fairly consistent picture of $\Teff$ and log(g) between the different observation epochs. Our statistical errors on each measured $\Teff$ and log(g) are very small: $\sim15$ K for \Teff and 0.01 dex for log(g). In reality, model uncertainties are expected to be larger so we report the weighted averages and adopt half a grid step as the error bars. In summary, we find $\Teff=4200\pm50$~K and log(g)$=4.40\pm0.25$, which agree within $1\sigma$ with literature values listed in Table~\ref{tab:prop}.

\onecolumngrid
\clearpage
\section{Fitting HIP~55507~A spectra with \texttt{petitRADTRANS}}\label{app:A_retrieval_model}

\begin{figure*}[h]
    \centering
\includegraphics[width=\linewidth]{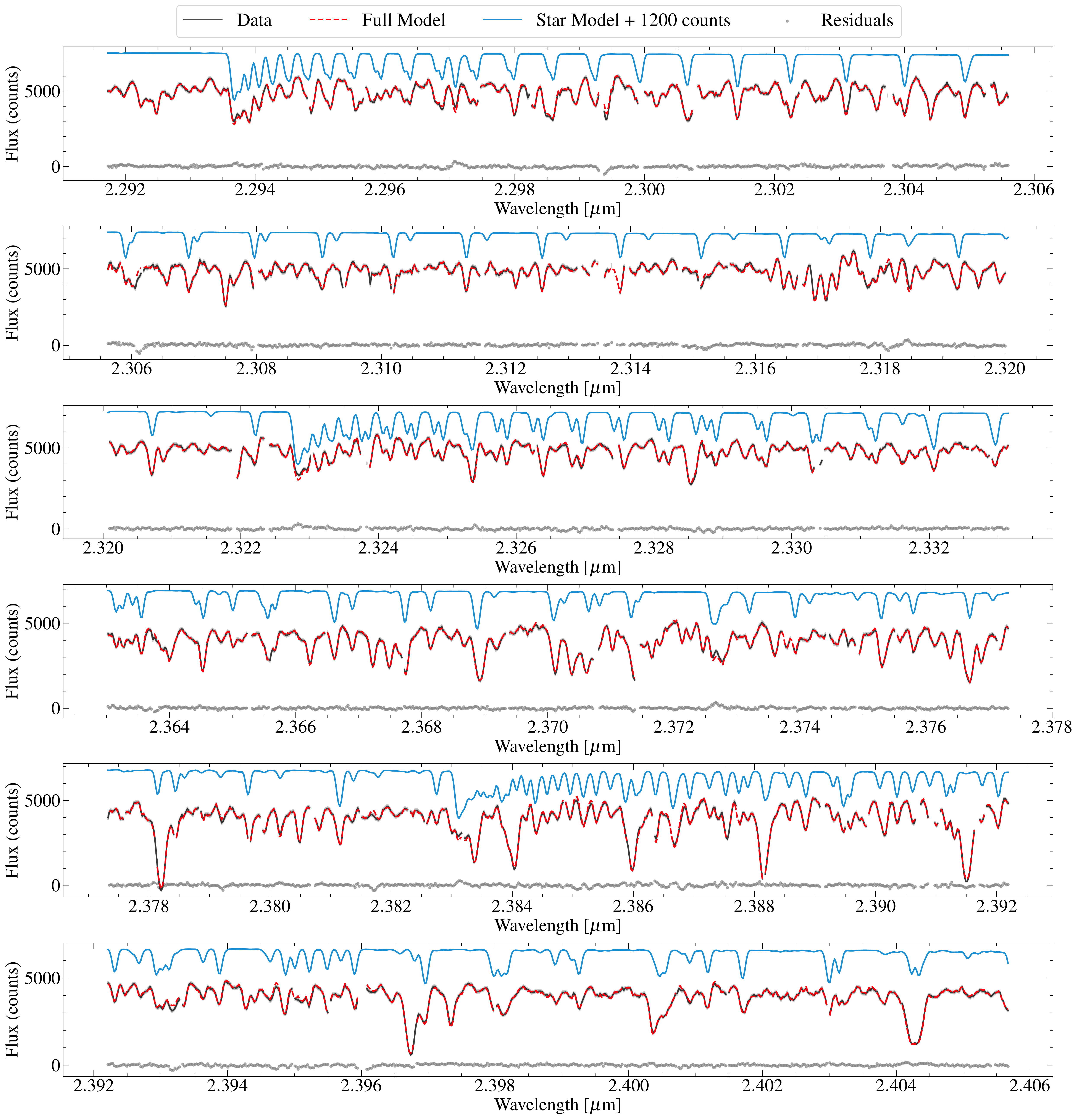}
    \caption{KPIC spectra for HIP~55507~A are plotted in black. We break up each spectral order into three panels. The full model ($FM_A$ in Eq.~\ref{eqn:star}) is in red, and includes the stellar model ($M_A$) from \texttt{petitRADTRANS} in blue and the telluric and instrumental response ($T$). The fringing model is also incorporated in the full model. The stellar model is offset by +1200 counts for clarity. The residuals are shown as gray points. CO lines dominate at these wavelengths for HIP~55507~A. We measure \co~and \coo~ from the spectrum.}
    \label{fig:A_retrieval_model}
\end{figure*}

\clearpage
\bibliography{main.bib}{}
\bibliographystyle{aasjournal}

\end{document}